\documentclass{emulateapj}
\usepackage{graphicx}
\usepackage{amsmath}

\def\araa{{\em ARAA}}
\def\apj{{\em ApJ}}
\def\apjs{{\em ApJS}}
\def\aap{{\em A\&A}}
\def\apss{{\em Astrophys.\ Space Science}}
\def\mnras{{\em MNRAS}}
\def\nat{{\em Nature}}
\def\pasj{{\em Pub.\ Astron.\ Soc.\ Japan}}

\newcommand{\ccm}{\,\mathrm{cm}^{-3}}
\newcommand{\htwo}{H$_2$ } 
\newcommand{\mhtwo}{\mathrm{H}_2\,} 
\newcommand{\lsob}{L_{\mathrm{s}}} 
\newcommand{\lsobcorr}{L_{\mathrm{s,corr}}} 
\newcommand{\lgnedin}{L_{\mathrm{g}}} 

\begin{document}

\title{A new approach to determine optically thick \htwo cooling and its effect on primordial star formation}

\author
{Tilman Hartwig, Paul C.\ Clark, Simon C.O.\ Glover, Ralf S.\ Klessen, Mei Sasaki}

\shorttitle{Improving Optically Thick \htwo Cooling}
\shortauthors{Hartwig et al.}

\affil{
Universit\"at Heidelberg, Zentrum f\"ur Astronomie, Institut f\"ur Theoretische Astrophysik, Albert--Ueberle--Str.\ 2, 69120 Heidelberg, Germany.
\break email: hartwig@iap.fr, p.clark@uni-heidelberg.de, glover@uni-heidelberg.de, klessen@uni-heidelberg.de, sasaki@stud.uni-heidelberg.de \\
}

\begin{abstract}
We present a new method for estimating the \htwo cooling rate in the optically thick regime in simulations of primordial star formation. Our new approach is based on the TreeCol algorithm, which projects matter distributions onto a spherical grid to create maps of column densities for each fluid element in the computational domain. We have improved this algorithm by using the relative gas velocities, to weight the individual matter contributions with the relative spectral line overlaps, in order to properly account for the Doppler effect.
We compare our new method to the widely used Sobolev approximation, which yields an estimate for the column density based on the local velocity gradient and the thermal velocity. This approach generally underestimates the photon escape probability, because it neglects the density gradient and the actual shape of the cloud.
We present a correction factor for the true line overlap in the Sobolev approximation and a new method based on local quantities, which fits the exact results reasonably well during the collapse of the cloud, with the error in the cooling rates always being less than $10\%$.
Analytical fitting formulae fail at determining the photon escape probability after formation of the first protostar (error of $\sim 40\%$) because they are based on the assumption of spherical symmetry and therefore break down once a protostellar accretion disc has formed.
Our method yields lower temperatures and hence promotes fragmentation for densities above $\sim 10^{10} \ccm$ at a distance of $\sim 200$\,AU from the first protostar. Since the overall accretion rates are hardly affected by the cooling implementation, we expect Pop III stars to have lower masses in our simulations, compared to the results of previous simulations that used the Sobolev approximation.

\end{abstract}

\maketitle

\section{INTRODUCTION}
\label{sec:intro}
The first stars in the Universe (so called Population III or Pop III stars) emerge several hundred million years after the Big Bang and dramatically change the physical conditions of their environment. The properties of Pop III stars have a fundamental influence on many subsequent physical processes, such as the synthesis of heavy elements, subsequent star and galaxy formation, or reionisation of the intergalactic gas, and it is therefore crucial to understand primordial star formation. Although a consistent and widely accepted formation scenario developed over the last few years \citep{g13,b13,g14b}, there are still a number of open questions, whose answers might modify this picture. A review of some of these these open questions is given by \citet{glover08}. In this paper, we focus on the determination of the optically thick \htwo cooling rate, because only little progress has been made on this 
topic since the late 80s in the context of star formation. Furthermore, \htwo emission is the dominant cooling process in this regime and hence a 
detailed understanding of its efficiency under different physical conditions is key to model Pop III star formation and gas dynamics in primordial halos.\\
Following the standard formation scenario, the first stars form at redshifts $z \simeq 20-30$ in dark matter halos that have total masses of $M \simeq 10^6 M_\sun$ and virial temperatures of around $1000$K \citep{bl04,g05,byhm09,cgkb11}.
Gas in the centre of these dark matter halos decouples and undergoes self--gravitating collapse \citep{yoha06}. The decoupled gas clouds have Jeans masses between $M_\mathrm{J}=200M_\sun$ and $M_\mathrm{J}=1000M_\sun$ \citep{abn00, abn02, yoha06, cgk08, mt08, tao09, cgsgkb11, hy13} and a critical number density of $n=10^4-10^5 \ccm$ when they decouple from their host halos \citep{abn00,cgkb11,b13}. During the collapse, gas cools mainly via \htwo rotational and vibrational line emission, with the \htwo fractional abundance being a few times $10^{-3}$ at the beginning of the collapse and becoming close to unity due to three--body \htwo formation above $10^8 \ccm$. Because these cooling processes can cool the gas very efficiently, the overall collapse proceeds almost isothermally until the gas becomes very optically thick at high densities and an adiabatic core forms. The exact determination of the optically thick cooling rate requires information about the velocity profile of the cloud, because 
relative velocities Doppler--shift 
the 
spectral lines 
and therefore 
increase the photon escape probability.
The shortcomings of commonly used methods for optically thick cooling lie mainly in the assumption of isotropy and in the dependence on local quantities.

Due to tidal forces and an initial angular momentum, the collapse does not proceed in a spherically symmetric manner, but rather leads to the formation of a rotationally supported disc around the first protostar. The disc generally extends out to $400-1000$AU, has a characteristic temperature of $1500-2000$K, becomes gravitationally unstable, and fragments into multiple parts \citep{sgb10, cgsgkb11, hoyy11, getal11, getal12, gsb13, g13, letal13}. The disc--like structure is a typical feature due to the inability of the halo to transfer angular momentum outwards quickly enough \citep{setal11}.

One of the main goals of simulations of primordial star formation is to establish the form of the Pop III Initial Mass Function (IMF). The mass of a Pop III star is the crucial parameter which defines its luminosity, temperature, spectrum, lifetime, final fate, and its metal yields. However, the significant variations and uncertainties in the expected mass ranges \citep{bkl01, op01, op03, oy03, jb06, mt08, oetal09, cgkb11, getal11, hoyy11, hetal14} reveal the lack of understanding of the primordial gas' fragmentation behaviour.
Since cooling is the key ingredient for disc instabilities and fragmentation, an accurate determination of optically thick \htwo cooling is of general interest for the determination of the primordial IMF.

Here, we analyse the dependence of fragmentation on two different cooling implementations: the commonly used Sobolev approximation and a more sophisticated, yet computationally expensive approach based on the TreeCol algorithm of \citet{cgk12}. We also present a new approximate method, which yields photon escape fractions with mean relative errors smaller than 10\%.

This paper is ordered as follows: In Section \ref{sec:methods}, we review the different approaches that are commonly used to model optically thick \htwo cooling. In Section \ref{sec:simulations}, we describe our numerical methodology. In Section \ref{sec:results}, we test the methods and present our results. We discuss these results in Section \ref{sec:discussion}, and conclude in Section \ref{sec:conclusion}.

\section{COOLING METHODS}
\label{sec:methods}
In this section, we present the commonly used photon escape probability approach for the determination of optically thick \htwo cooling rates and several methods of calculating this probability. Moreover, we derive a correction factor for the Sobolev approximation to account for the overlap of spectral lines.

\subsection{Optically Thick Cooling}
Since \htwo is a symmetric, diatomic molecule, it has no permanent dipole moment and can therefore only radiate via ro--vibrational transitions \citep{lpf99}. In the optically thin regime $(n \leq 10^9 \ccm)$ we expect all radiation to escape the cloud freely and the cooling rate can be calculated fairly accurately as a function of the local density, temperature and chemical composition of the gas. A number of different parametrisations of the \htwo cooling rate in terms of these quantities are available in the literature. In this work, we use the rates given in \citet{ga08}.
In the optically thin case we assume that photons, which are emitted in the ro--vibrational transitions, can escape the cloud without being scattered or absorbed and therefore transport thermal energy outwards efficiently. In the optically thick case, however, scattering and absorption events might capture the photons and hence decrease the overall cooling efficiency. One generally assumes complete redistribution of the frequency between these scattering events and uses the ``escape probability method'' in order to solve this problem. Most important for our further discussion are the applications of the escape probability method by \citet{yoha06} and \citet{cgkb11} to the case of optically thick cooling in primordial gas. The cooling rate in an optically thick medium is given by
\begin{equation}
 \Lambda _{\mhtwo,\mathrm{thick}} = \sum _{l,u} E_{lu}A_{lu} \beta _{\mathrm{esc},lu} n_u,
\end{equation}
where $E_{lu} = h \nu _{lu}$ is the energy separation between the lower level $l$ and upper level $u$, $A_{lu}$ is the spontaneous radiative transition rate for transitions between $u$ and $l$, $\beta _{\mathrm{esc},lu}$ is the probability for an emitted line photon to escape without absorption, and $n_u$ is the population density of hydrogen molecules in the upper level. Following \citet{yoha06}, we assume all energy levels to be populated according to local thermodynamic equilibrium and we consider rotational levels from $J=0$ to $20$ and vibrational levels $\nu = 0,1,2$. At the temperatures of interest, the contribution of other levels will be negligibly small ($<10^{-3}$). We take values for the level energies from the compilation made available by P. G. Martin on his website\footnote{\url{http://www.cita.utoronto.ca/~pgmartin/h2.html}} and the radiative transitions rates from \citet{wsd98}. Since all other quantities are known for typical conditions in primordial gas, the remaining task is 
the determination of the photon escape probability.

\subsection{Sobolev Approximation}
Based on the work by \citet{z34}, \citet{s47} derived the escape probability for constant velocity gradients in his study of expanding envelopes (for an English translation see \citealt{s60}). This approximation has been reviewed during the last few decades \citep{c70,hr82,yoha06} and is widely used in simulations of primordial star formation \citep{yoha06, yoh08, tetal11, cgsgkb11, cgkb11, getal11, whb11, getal12, hy13, sgkbl13, sb13, gsb13, hetal14, g14a, sb14}. Following the derivation by \citet{e96}, the photon escape probability is given by
\begin{equation}
 \beta _{\mathrm{esc},lu} = \frac{1-e^{-\tau _{lu}}}{\tau _{lu}},
\end{equation}
where $\tau _{lu}$ is the opacity at the line centre. The absorption coefficient for a transition from the lower to the upper level is given by
\begin{equation}
 \alpha _{lu} = \frac{E_{lu}}{4 \pi}n_l B_{lu} \left[ 1- \exp \left( \frac{-E_{lu}}{k_B T} \right) \right] \phi (\nu),
\label{eq_alpha}
\end{equation}
where $B_{lu}$ is the Einstein coefficient for absorption and $\phi (\nu)$ is the line profile function. The opacity can be written as
\begin{equation}
 \tau _{lu} = \alpha _{lu} L_\mathrm{char},
\end{equation}
where $L_\mathrm{char}$ is a characteristic length scale \citep{c70, gk74, ddc75, sp05}. According to equation (\ref{eq_alpha}), $\alpha _{lu}/n_{\mhtwo}$ is only a function of temperature and, following \citet{cgkb11}, we can express the optical depth by
\begin{equation}
 \tau _{lu} = \left( \frac{\alpha _{lu}}{n_{\mhtwo}} \right) n_{\mhtwo} L_\mathrm{char},
\end{equation}
where
\begin{equation}
 n_{\mhtwo} L_\mathrm{char} = N_{\mhtwo,\mathrm{eff}}
\label{eq_effN}
\end{equation}
defines an effective column density. Hence, the last task for the determination of optically thick cooling is the calculation of the characteristic length and the associated effective column density. Knowing these quantities, one can generally determine the angle--dependent escape probability and afterwards average it over all lines of sight and all relevant ro--vibrational lines. Since \htwo cooling is due to a number of lines without a single dominant line \citep{hrcf02}, we can directly average the escape probability over these lines. Furthermore, one often assumes spherical symmetry, whereas \citet{yoha06} proposed an average over three orthogonal directions
\begin{equation}
 \beta = \frac{\beta _x + \beta _y + \beta _z}{3}.
\end{equation}
The averaged escape probability finally relates the optically thin and optically thick cooling rate by
\begin{equation}
 \Lambda _{\mhtwo,\mathrm{thick}} = \beta \Lambda _{\mhtwo,\mathrm{thin}}
 \label{eq_thickcooling}
\end{equation}
and is therefore also known as the ``opacity correction''. In order to determine this escape probability, we have to understand the dynamics of the cloud. If the photon is emitted in the centre of the cloud and an envelope of gas is moving towards it with a constant radial velocity gradient $\mathrm{d} v_\mathrm{r} / \mathrm{d} r$, then the photon observes the spectral lines of the envelope to be Doppler--shifted with respect to its rest frame. According to \citet{s47}, a photon is not absorbed and can escape freely, if the spectral lines of a possibly absorbing \htwo molecule are shifted by more than one thermal line width. The line width of thermal line broadening is given by
\begin{equation}
 \Delta \nu _{\mathrm{th}} = \nu _0 \frac{v_{\mathrm{th}}}{c} = \frac{\nu _0}{c} \sqrt{\frac{2k_BT}{m_{\mathrm{H}_2}}},
\end{equation}
where $\nu _0$ is the central frequency of the line, $v_{\mathrm{th}}$ is the thermal velocity of molecular hydrogen, $T$ is the temperature, and $m_\mathrm{H}$ is the mass of a hydrogen atom. Using this, we can determine the characteristic distance $L_\mathrm{char}$ beyond which the Sobolev criterion is fulfilled. This length scale is typically known as the Sobolev length
\begin{equation}
 \lsob = \frac{v_\mathrm{th}}{|\mathrm{d}v_\mathrm{r} / \mathrm{d}r|}.
 \label{eq_sob_length}
\end{equation}
Phrased differently, all relevant matter that might reabsorb a photon is within its Sobolev length. Assuming a constant density within this Sobolev length, the effective column density can be determined by
\begin{equation}
 N_{\mhtwo,\mathrm{eff}} = n_{\mhtwo} \lsob.
\label{eq_Lsob}
\end{equation}
In order to capture the three--dimensional dynamics of the collapse, one normally uses
\begin{equation}
 \lsob = \frac{v_\mathrm{th}}{|\mathbf \nabla \cdot {\mathbf v}|}
\end{equation}
for the determination of the Sobolev length \citep{nk93}. However, a fundamental problem of the Sobolev approximation was already mentioned by several authors: both the velocity gradient and the number density have to be constant within one Sobolev length \citep{l71, bgno80, hr92, nk93, whb11}. As we will see below, this assumption is generally not valid.

\subsection{Correction of Sobolev Approximation}
\begin{figure}[t]
\centering
\includegraphics{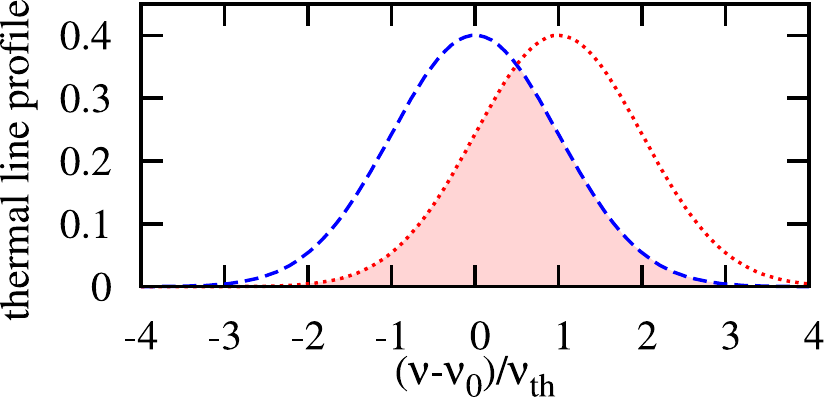}
\caption{\label{fig:overlap} {Normalised thermal line profiles as a function of frequency in units of the thermal line width. The right profile is shifted by one thermal line width with respect to the left profile. The overlapping area shows a relative overlap of 62\%, which should be negligible according to \cite{s47}.}}
\end{figure}

For simplicity, \citet{z34} assumed the absorption coefficient $\alpha _{lu}$ to be zero outside the interval $[\nu _0 - \Delta \nu _{\mathrm{th}},\nu _0 + \Delta \nu _{\mathrm{th}}]$ and \citet{s47} used the same simplification.
Following this approach, a photon can escape freely from the optically thick gas after one Sobolev length $\lsob$, because it will not be reabsorbed thereafter. The actual absorption probability however, is related to the true overlap of spectral lines. In the present context, the shape of the \htwo lines is dominated by thermal broadening and can therefore be described by a normalised Gaussian profile. If one line with central frequency $\nu _0$ is Doppler--shifted to the frequency $\nu$, the relative line displacement in units of the thermal gas velocity between these two cases is given by
\begin{equation}
x=\frac{\nu - \nu _0}{\Delta \nu _{\mathrm{th}}}.
\label{eq_linedisplace}
\end{equation}
Accordingly, the relative overlap of these two line profiles can be calculated by
\begin{equation}
o(x)=\int _{- \infty} ^{x/2} \frac{2}{\sqrt{2 \pi}} \left( e^{- \nu ^2/2} - e^ {- (\nu -x)^2/2} \right) \rm d \nu.
\label{eq_overlap}
\end{equation}
The overlap of spectral lines for the special case $x=1$ is illustrated in Figure \ref{fig:overlap}.
This displacement represents the case after one Sobolev length, though the relative overlap is still $62\%$. While the original Sobolev approximation ignores all possible absorption events beyond this point, there is a non--negligible absorption possibility beyond one Sobolev length and we therefore need to introduce a correction term in order to account for the additional matter. Even after three Sobolev lengths the relative overlap, for example, is still 13\% and reabsorption might be possible.

In order to find a proper correction for this overlap, we start from the basic definition of the column density. However, we are not interested in the total column density but rather in the effective column density which includes only the gas that could be relevant for the reabsorption of escaping \htwo line photons. The Sobolev approximation gives a very simple answer to the question of which gas we have to include in the effective column density, namely all gas within one Sobolev length. Expressed in terms of line overlap we write this as
\begin{equation}
   N_{\mathrm{s}}= \underbrace{\int _0 ^{\lsob} 1 \cdot n_{\mhtwo} \mathrm{d} r}_{100\% \, \mathrm{overlap}} + \underbrace{\int _{\lsob} ^{\infty} 0 \cdot n_{\mhtwo} \mathrm{d} r}_{0\% \, \mathrm{overlap}} = n_{\mhtwo} \lsob ,
\label{sob_overlap}
\end{equation}
where a constant number density of molecular hydrogen is assumed in the last step. Thus, \citet{s47} overestimated the overlap and the matter contribution within one Sobolev length but neglected all matter contributions beyond this point. Since we want to account for the true overlap of spectral lines, we introduce the relative line overlap $o(x)$ as a weighting function into the determination of the effective column density,
\begin{equation}
 N_{\mathrm{s,corr}} = \int _0 ^{\infty} o(x) \cdot n_{\mhtwo} \mathrm{d} r.
\end{equation}
The remaining problem is to relate the line displacement $x$ to the distance $r$ along the line of sight. Assuming a constant velocity gradient $\mathrm d v_{\mathrm{r}}/\mathrm{d} r$, we can rewrite equation (\ref{eq_sob_length}) into
\begin{equation}
 \mathrm{d} r = \frac{\lsob}{v_{\mathrm{th}}} \mathrm d v_{\mathrm{r}}
\end{equation}
and transform the radial integration into an integration over the relative line displacement
\begin{equation}
  N_{\mathrm{s,corr}} = n_{\mhtwo} \lsob \int _0 ^{\infty} o(x)  \mathrm d x,
\end{equation}
which finally yields
\begin{equation}
  N_{\mathrm{s,corr}} \simeq 1.694 N_{\mathrm{sob}}.
\label{eq:correction_factor}
\end{equation}
In other words, the relevant column density for the determination of the escape probability of line photons is about 1.7 times higher than originally assumed by Sobolev.
Although we correct for the line overlap, we should still keep in mind that this derivation implies several assumptions like constant velocity gradient and constant density within the Sobolev length.
Also some gas further away may be absorbing, because of similar velocities. In particular in turbulent clouds this may become relevant. This is related to the question of how ``real'' clumps identified in position--position--velocity space are in position--position--position space (see e.g. \citealt{bm02} and \citealt{bosgg13}).

Molecular hydrogen has more than two hundred spectral lines in the range $1.0 \mu \mathrm m \leq \lambda \leq 32 \mu \mathrm m$ which could be relevant for cooling primordial gas \citep{rhfc02}.
Although these lines appear to be very close to each other, the possibility that a photon is emitted in one line with a certain frequency and absorbed by another line which is Doppler--shifted into the emitting frequency is negligibly small ($\sim 10^{-6}$). Consequently, this effect can be ignored.

\subsection{Further Cooling Approaches}
Besides the commonly used Sobolev approximation there are several other approaches, which we present in the following subsections.

\subsubsection{Gnedin Approximation}
\citet{gtk09} model molecular hydrogen and star formation in cosmological simulations. They determine \htwo column densities for the self--shielding of \htwo against Lyman--Werner photons by using a ``Sobolev--like'' approximation. They define a characteristic length scale
\begin{equation}
 \lgnedin = \frac{n_{\mhtwo}}{|\mathbf{\nabla} n_{\mhtwo}|}
\end{equation}
based on the number density and its gradient. The column density is then simply given by
\begin{equation}
 N_{\mathrm{g}} = n_{\mhtwo} \lgnedin .
\end{equation}
This approach accounts for the density gradient in the gas distribution. In the following, we will label this method as ``Gnedin''. They claim that this approximation provides a very good estimate for the column density in the range $3 \times 10^{20} \mathrm{cm}^{-2} < N_{\mathrm{HI}}+2N_{\mhtwo} < 3 \times 10^{23} \mathrm{cm}^{-2}$. Although we are dealing with star formation on much smaller scales than they do, their method seems to be a reasonable, scale--invariant approach for the determination of effective column densities. Furthermore, this method depends only on local quantities and is easy to implement in different codes. Nevertheless, this method includes no information about the velocity profile of the cloud and therefore neglects the enhanced photon escape probability due to the Doppler--shifting of lines.

\subsubsection{Analytic Fit Functions}
We are interested in the column densities in order to determine the photon escape probabilities for \htwo line cooling (equation \ref{eq_thickcooling}). Besides the previously presented methods, there are two analytical fitting functions that directly relate a given number density of the gas to the photon escape probability. The first fitting function
\begin{equation}
 \beta = \min \left[ 1, (n/n_\mathrm{RA})^{-b_{\mathrm{RA}}} \right]
\label{RA_fit_formula}
\end{equation}
with $n_\mathrm{RA} = 8 \times 10^9 \ccm$ and $b_{\mathrm{RA}}=0.45$ was proposed by \citet{ra04} and has been applied in several (mainly grid--based) simulations \citep{on06, tetal11, hy13, g14a}. This formula was obtained from the detailed one--dimensional calculation by \citet{rhfc02}.
A second method was proposed by \citet{gsb13} who study the chemo--thermal instability in primordial star--forming clouds. The idea follows the approach by \citet{ra04} but with a smooth transition and therefore a continuous derivative towards the optically thin regime. The formula is given by
\begin{equation}
 \beta = \begin{cases}
          \frac{(1+b_{\mathrm{G}})x}{x^{(1+b_{\mathrm{G}})}+b_{\mathrm{G}}} &\mathrm{for}\, x \geq 1 \\
          1 &\mathrm{for}\, x < 1
         \end{cases}
\label{Greif_fit_formula}
\end{equation}
where $x=n/n_\mathrm{G}$, $n_\mathrm{G} = 4 \times 10^9 \ccm$, and $b_{\mathrm{G}}=0.45$. These fits are the easiest and most direct way to determine the photon escape probability, but their use does not account for any information about the temperature, velocity or density profiles. Recently, \citet{g14a} investigated the collapse of primordial gas using a multi--line, multi--frequency raytracing scheme in order to accurately model the transfer of \htwo line emission. Although he provides a new fit formula, we will not include this in our analysis, since it is very similar to the original fit function by \citet{ra04}.

\subsubsection{Reciprocal Approach}
In order to find a method for the determination of optically thick cooling that only depends on local quantities of the collapse and that is easy to implement, but nevertheless captures the dynamics of the cloud, we propose a combination of the Sobolev and Gnedin approximation. The corrected Sobolev approximation takes the line overlap into account but neglects the decreasing density. On the other hand, the Gnedin approximation takes the decreasing number density into account but neglects the Doppler--shifting of lines. Since each method on its own generally overestimates column densities, the general idea behind this new approach is to combine these two methods in order to overcome their individual shortcomings. The reciprocal sum
\begin{equation}
 \frac{1}{L_{\mathrm{rec}}} = \frac{1}{\lgnedin} + \frac{1}{\lsobcorr}
\end{equation}
of the two characteristic lengths provides the right behaviour. If both lengths are of the same order ($L_1 \simeq L_2$), the result should be smaller than both lengths ($L_{\mathrm{rec}} < L_1 \simeq L_2$), since each length individually overestimates the column density. Whereas if one length is significantly smaller than the other length $L_1 \ll L_2$, the result should be equal to the smaller one ($L_{\mathrm{rec}} \simeq L_1$), because beyond this smaller distance the photons can escape freely anyway. Following this method, the number density is simply given by
\begin{equation}
 N_{\mathrm{rec}} = n_{\mhtwo} L_{\mathrm{rec}}.
\end{equation}

\section{NUMERICAL METHODS}
\label{sec:simulations}
Here, we describe the implementation of TreeCol and three criteria to define the effective column density. Furthermore, we present our simulations and the initial conditions.

\subsection{Effective Column Densities with \sc TreeCol}
\label{sec:coolingmethods}
The main problem of the Sobolev approximation is the dependence on local quantities and therefore the neglect of all information about density gradients, velocity profile and the actual shape of the cloud. Generally, the most exact way to determine effective column densities is to sum up all relevant mass along all possible lines of sight. In order to avoid this extremely high computational effort, \citet{cgk12} designed the TreeCol algorithm which determines column densities based on a tree structure, used by many gravitational $N$--body solvers. TreeCol uses a spherical pixelation with diamond--shaped pixels based on HEALPix \citep{getal05}. During the walk of the tree, all relevant data for the column density map are collected and projected onto a spherical grid.
Since the data are already stored in a tree, TreeCol can use this information and therefore scales as $N\log N$ with the number of cells or particles $N$. However, in our simulations, the usage of TreeCol slows down the simulation by a factor of about five with respect to a run without TreeCol. This slowdown is mainly related to the evaluation of several inverse trigonometric functions. Although we use this method in an SPH--based simulation, the only requirement for its implementation is the clustering of matter in a tree--like structure, as indeed, the TreeCol method has already been implemented in the Arepo moving mesh code \citep{setal14} and the FLASH AMR code (W\"unsch et al., in prep.).\\
TreeCol overcomes several shortcomings of the Sobolev approximation. First of all, we can use it for any density distribution because we directly sum up the individual mass contributions. Furthermore, we can use the velocity information of the tree nodes and do not have to assume a constant velocity gradient. Additionally, we account for the actual spatial matter distribution and do not have to stick to a rough one dimensional approximation.
While the original TreeCol algorithm computes the \htwo column densities of the entire cloud, in what follows, we will present three improved versions that account for the Doppler--shifting of the line by only including mass that lies within the appropriate velocity range.
\begin{figure}[t]
\centering
\includegraphics[width=0.45\textwidth]{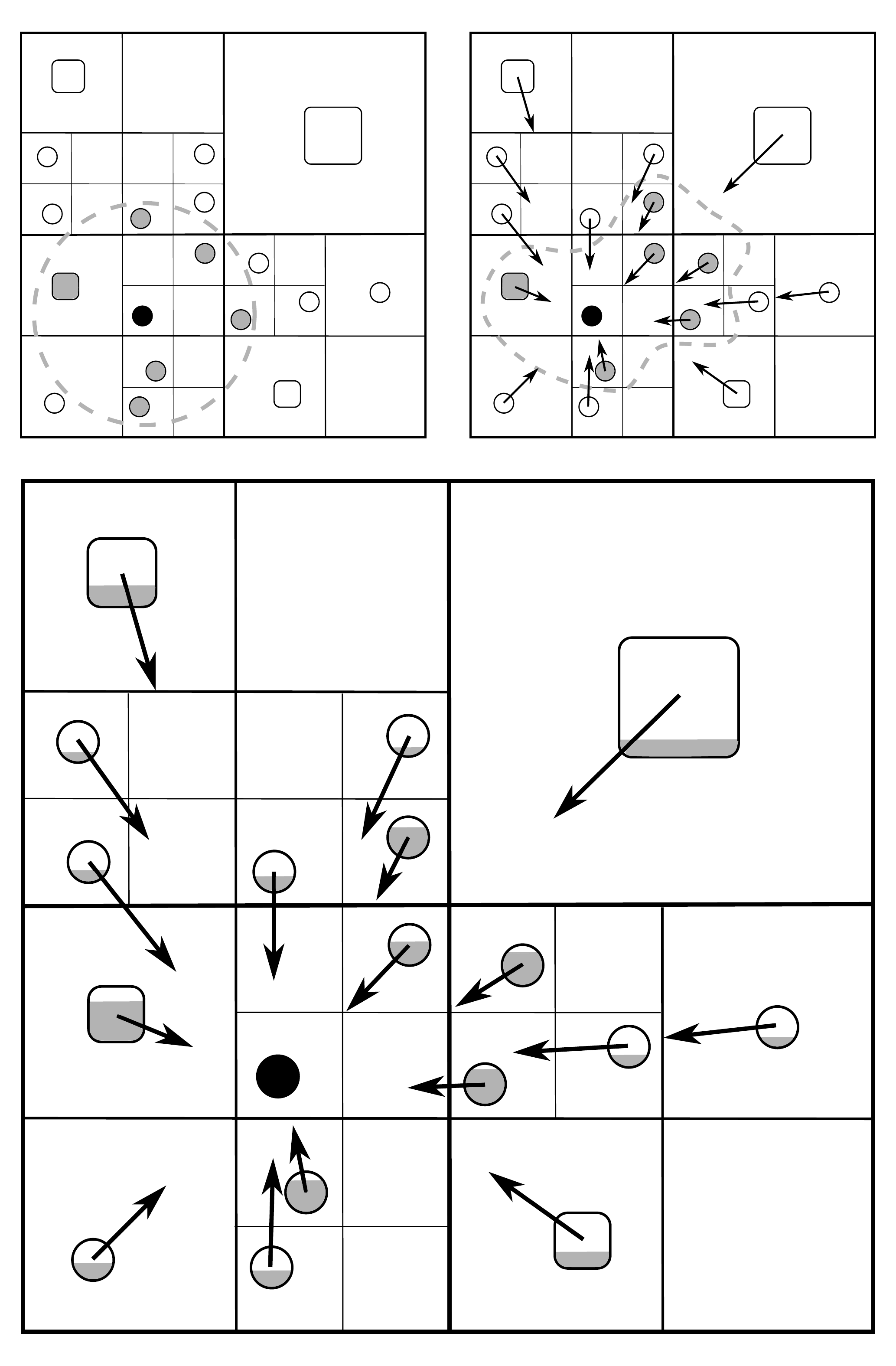} 
\caption{Two--dimensional illustration of three different approaches for the determination of effective column densities. \textit{Top left}: original Sobolev method without TreeCol. \textit{Top right}: Sobolev--like method implemented in TreeCol. \textit{Bottom}: Lookup method implemented in TreeCol. The circles represent individual mass contributions from cells or particles and the squares represent all mass in this tree node, which are clustered as seen by the black particle. The arrows indicate the relative velocities with respect to the black target particle, for which the column density should be determined.
The original, isotropic Sobolev method defines a characteristic length scale, within which all matter contributes to the effective column density.
In order to affect the effective column densities of the target particle in the Sobolev--like implementation, the relative velocities of the tree nodes have to be smaller than the thermal velocity $v_\mathrm{th}$ of the target particle.
In the lookup implementation, the matter contributions to the column density are weighted by the relative overlap of spectral lines, which in turn depends on the relative velocities. This weighting is illustrated by the partial filling.}
\label{fig:scheme}
\end{figure}

\subsubsection{Sobolev--Like}
The Sobolev approximation assumes that all relevant mass for the column density is located within one Sobolev length. Translated into velocities, we should only include particles or tree nodes whose relative velocity is smaller than the thermal velocity. Thus, we modify TreeCol in order to include only the mass contributions of nodes that fulfil this criterion. A schematic illustration of this approach is given in Figure \ref{fig:scheme}. From there we already see that the relevant volume around a particle does not necessarily have to be spherical, but usually follows the dynamic shape of the cloud.

\subsubsection{Corrected Sobolev}
Following equation (\ref{eq:correction_factor}), the next possible approach is to include all tree nodes whose relative velocities fulfil the criterion $|v_\mathrm{r}| < 1.694 v_\mathrm{th}$. Using this criterion, we take into account the non--negligible overlap beyond one Sobolev length. However, the individual velocity contributions are still not taken into account correctly, and we will focus on the following, more sophisticated approach for further studies.

\subsubsection{Lookup Method}
\label{sec:lookupmethod}
The above methods only distinguish whether matter contributes to the effective column density or not. However, the Doppler--shifting of spectral lines is a smooth process and so is the matter contribution, which should be included in a proper treatment of line cooling.
The overlap of spectral lines and therefore the relevance for the effective column density depends only on the relative velocity. Since we know this information for the individual tree nodes, we weight their contributions to the column density map with their relative spectral line overlap (equation \ref{eq_overlap}). A schematic illustration of this approach is given in Figure \ref{fig:scheme}. This method does not rely on any assumptions (like a constant velocity gradient or number density), but captures the complete three dimensional collapse of the cloud and takes care of the true line overlaps. This is the most exact method and it will serve as a reference for our analysis. For simplicity, we will refer to this method simply as ``TreeCol'', but actually mean the lookup method implemented in TreeCol --- if not explicitly stated otherwise.

\subsection{Simulations}
In this section, we describe the general properties of our code, discuss the implementation of the TreeCol method, comment on its computational effort, and present our initial conditions.

\subsubsection{Methods}
We use a modified version of the cosmological simulation code Gadget--2. The original code was written by \citet{s05} in order to simulate structure formation by means of smoothed particle hydrodynamics (SPH).
For the nonlinear collapse, we follow the chemical and thermal evolution of the gas cloud, solve the rate equations and couple the relevant heating and cooling terms to the hydrodynamic equations. For this purpose, we use a chemical network and cooling functions developed by Glover and collaborators (see \citealt{gj07, ga08, cgsgkb11} and \citealt{cgkb11} for full details).
The primordial chemistry network includes H, D, He, H$_2$, H$^+$, H$^-$, D$^+$, H$_2^+$, HD, He$^+$, He$^{++}$ and e$^-$. The rate equations between these species are solved self--consistently for every time step. For the three-body \htwo formation reaction
\begin{equation}
\mathrm{H} + \mathrm{H} + \mathrm{H} \rightarrow \mhtwo + \mathrm{H}
\end{equation}
we adopt the rate coefficient of \citet{glov08}, derived by applying the principle of thermal balance to the collisional dissociation rate of \citet{msm96}.
The time scales associated with chemical heating and cooling can become extremely short when species come close to chemical equilibrium. Since the particle time step depends on the thermal timescale, this can result in some particles having very short timescales once the gas density exceeds $\sim 10^{17} \ccm$. At this point it can become computationally prohibitive to run the simulation further.

Besides \htwo line cooling, which is the most relevant process for this analysis, we also track the contribution by several other heating and cooling processes, such as electronic excitation of H and He, recombination, photodissociation, HD line cooling, Compton cooling, and bremsstrahlung \citep[see][for a more detailed description]{ra04, gj07, cgkb11}. For the collisional induced emission (CIE) cooling, we follow the implementation by \citet{cgsgkb11}, which is based on studies by \citet{rhfc02,ra04}, and \citet{yoh08}. Above densities of $n_\mathrm{CIE} \approx 10^{14}\ccm$, CIE cooling from \htwo becomes more efficient than \htwo line cooling, but the gas is most susceptible to fragmentation at densities below $n_\mathrm{CIE}$ (see Section \ref{sec:frag2}). Consequently, the implementation of CIE cooling is unlikely to significantly affect the conclusions of our study.

Furthermore, we implement a simple feedback model for the accretion heating by assuming that the accreting protostars produce an accretion luminosity of $L_\mathrm{acc} = G M_{*} \dot{M}/R_{*}$, where $\dot{M}$ is the mass accretion rate and $M_*$ and $R_*$ are the mass and radius of the protostars, respectively. We use the models by \citet{setal11} for the protostellar radius and assume a constant accretion rate of $\dot{M}=10^{-2} M_\sun \mathrm{yr}^{-1}$, which is consistent with \citet{cgsgkb11}, \citet{hy13}, and with the actual accretion rates in our model. The heating rate is then given by
\begin{equation}
 \Gamma _\mathrm{acc} = \rho \kappa _P \left( \frac{L_\mathrm{acc}}{4 \pi r^2} \right),
\end{equation}
where $\rho$ is the gas density, $\kappa _P$ is the Planck mean opacity and $r$ is the distance from the source.

In order to minimise computational effort in the high--density regime, we use sink particles, based on the implementation by \citet{jetal05}, which was already used before \citep[e.g.][]{bb82, b87, b89, bbp95}. Above a certain density threshold $n_{\mathrm{crit}}$ all SPH particles are merged into one single sink particle, which now contains all mass and momenta of the merged particles. This approach conserves mass and momentum, avoids small dynamical and chemical time steps, and we can identify the protostars by the newly formed sink particles. The critical density threshold is chosen so that the local Jeans mass is resolved by at least 100 SPH particles, following the resolution criterion by \citet{bb97}.\\
Nevertheless, one should keep in mind that sink particles are not physical entities in their own right, but rather computationally motivated and consequently may cause problems by introducing a discontinuity in the mass and number of particles as well as a lack of pressure forces at the accretion radius. Strictly speaking, they also violate the hydrodynamic equations because the accretion onto the sink particle happens instantaneously \citep{getal12}. In order to guarantee that the formation of a sink particle actually represents the local collapse to a protostar, we introduce several formation criteria. The gas clump converted into a sink particle must exceed a certain density threshold, it must be gravitationally unstable, at the centre of a locally convergent flow, and it must have a certain distance to preexisting sink particles (see e.g.\ \citealt{fbck10}).

Another crucial value is the accretion radius $r_\mathrm{acc}$, because a too large accretion radius might artificially influence the fragmentation behaviour \citep{cgsgkb11, getal11, md13}. On the other hand, \citet{getal12} and \citet{md13} state that fragmentation does not occur for densities above $n \simeq 10^{17} \ccm$, because there are no more efficient chemical or radiative cooling mechanisms. \citet{setal11}, who additionally include the effect of heating by accretion feedback, find this value to be $n \simeq 10^{15} \ccm$. Phrased differently, the choice of the accretion radius might have an influence on the fragmentation behaviour but as long as the critical density is $n_{\mathrm{crit}} \gtrsim 10^{16} \ccm$, we should capture all fragmentation of the cloud.
We set the critical density to $n_\mathrm{crit} = n_\mathrm{res}=3.42 \times 10^{15} \ccm$ in order to resolve the Jeans mass throughout the simulation. The Jeans length is $\lambda _J =0.3$AU under these conditions. Such a small accretion radius might lead to tiny dynamical time steps in the vicinity of sink particles and hence increase the computational effort. Therefore, we set the accretion radius to $r_\mathrm{acc} =10$AU, which clearly fulfils the resolution criterion, but may suppress small--scale fragmentation.

\subsubsection{Implementation of Cooling Approaches}
The general idea of the different cooling approaches has already been discussed in section \ref{sec:coolingmethods}. Here, we address the actual implementation of these approaches into the code.\\
The optically thick cooling rate $\Lambda _{\mhtwo,\mathrm{thick}}$ is given by equation (\ref{eq_thickcooling}), where the photon escape probability $\beta$ can be expressed as a function of the column density divided by the thermal velocity of the \htwo molecules ($N/v_\mathrm{th}$) and the temperature $T$. These values are stored in a lookup table for $31.6\ \mathrm{K} \leq T \leq 31600\ \mathrm{K}$ and $10^{17} \mathrm{s}\, \ccm \leq N/v_\mathrm{th} \leq 10^{27} \mathrm{s}\, \ccm$.

For the TreeCol--based determination of the effective column densities we create another lookup table, which relates the relative velocities in units of the thermal velocities ($v_\mathrm{r}/v_\mathrm{th}$) to an overlap of spectral lines. Since the relative velocities are distributed roughly equally throughout the simulation, we create this lookup table with linear steps in velocity space. For each node that might contribute to the effective column density we first check if
\begin{equation}
\left( \frac{v_\mathrm{r}}{v_{\mathrm{th}}} \right) ^2 \leq 43.3
\end{equation}
because otherwise the line overlap is smaller than $10^{-3}$ and can be neglected anyway. The computational effort for the lookup of relative overlaps is negligibly small compared to the computational cost of TreeCol itself.\\

\subsubsection{Initial Conditions}
\label{sec:IC}
The initial conditions of our simulations presented here are generated in the following way:
First, we ran collisionless $N$--body simulations with 1 Mpc/$h$ comoving in length with uniform mass resolution to capture collapsing dark matter halos starting at a redshift $z$=100. These collisionless simulations were executed with Gadget--3 \citep{setal05} and we employ cosmological parameters consistent with the WMAP--7 measurements ($\Omega_m = 0.271$, $\Omega_{\Lambda} = 0.729$, $\sigma_8 = 0.809$, $h = 0.703$, \citealt{ketal11}). We would not expect our results to differ significantly if we were to use the cosmological parameters measured by Planck \citep{Pla13}. This choice of parameters is for consistency with our previous works \citep{scskg14}. Since we ensure that the parameters for the tree force calculation are set to the same value in the Gadget--3 and Arepo simulations, described later in this section, this is practically equivalent to using Arepo for the N--body simulation. The reason we used Gadget--3 to evolve dark matter only initial conditions is that, in principle, it allows us 
to make use of more physical merger histories of dark matter halos when picking a specific dark matter halo to resimulate. Currently, this is only analysed with ease from Gadget--3 snapshots but not from Arepo snapshots.
From this parent simulation, we
select four dark matter halos, which are individually shifted to the centre of the simulation box. The regions further away from each specific halo are represented by dark matter particles of proceedingly larger mass and lower resolution. The parent simulation contains $256^3$ (halo 1) and $512^3$ (halo 2-4) particles. At the regions of interest the mass resolution of the dark matter particles is improved by $4^3$. The dark matter mass is 7.3 $\rm M_{\odot}$ for the $512^3$ runs.  

From this newly generated initial condition with improved dark matter resolution, we start a hydrodynamical simulation with the moving--mesh code Arepo \citep{s10}.
At the start--up of our hydrodynamic simulations, the code generates gas cells from dark matter only initial conditions. In order to numerically follow the dynamics of pristine gas in a cosmological context, we adopt an on--the--fly mass refinement scheme, which ensures that the Jeans length is resolved by at least 64 cells at all times. In order to deal with memory consumption and small time steps, 
we stop our simulation when the highest density in the simulation is $\sim$ $10^7 \rm cm^{-3}$. 

We cut out spheres with a radius of $R=0.5$\, pc, store only gas cells, and continue the simulation forcing further mass refinement in a non--cosmological setup (256 cells per Jeans length). This is to ensure that the Jeans mass at $\rm n \sim 10^{17} \rm cm^{-3}$ is resolved with $\sim$ 100 cells without further refinement in the Gadget--2 simulation.

The resulting gas clouds serve as initial conditions for the high--resolution runs, which are performed with Gadget. These clouds have average masses of $\sim 1.0 \times 10^3 M_\sun$, temperatures of $\sim 400$\,K, and a mean \htwo abundance of $x_{\mathrm{H}_2} = 1.1 \times 10^{-3}$. The average particle number is $\sim 2 \times 10^7$ per simulation, which yields a mass resolution of $10^{-3} M_\sun$ in the central region. These final runs are performed in Gadget--2, to make use of the already implemented primordial chemistry.

An additional external pressure term is added in order to compensate for the missing gas contribution from the surrounding halo \citep{b90, cgkb11}. In order not to artificially squeeze the cloud, we set the external pressure to the smallest occurring pressure in the outer 10\% of the cloud.

\section{RESULTS}
\label{sec:results}
In this section, we first test the validity of our new method in a simple, spherically symmetric test scenario. Then, we compare the TreeCol--based cooling approach to the local, isotropic cooling implementations for the cosmological halos.

\subsection{Test Scenario}
\begin{figure}[t]
\centering
\includegraphics[angle=-90]{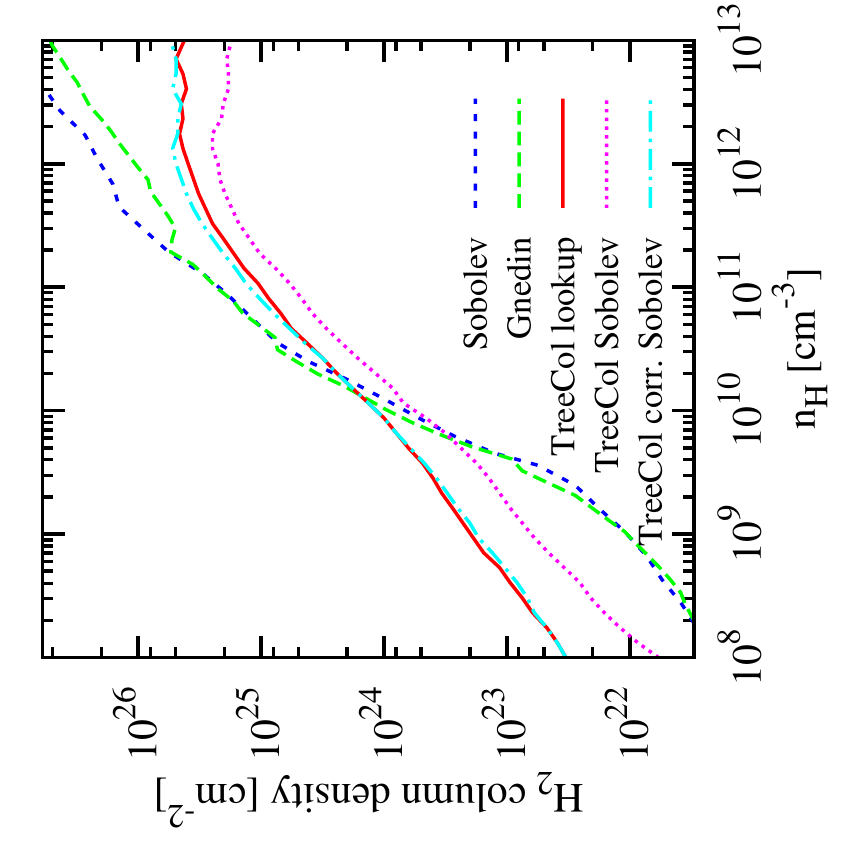}
\caption{Effective \htwo column density as a function of density for different cooling approaches. Here and in the following figures (if not stated otherwise), the plotted values represent the means of the density-binned SPH particles. Staring from spherically symmetric initial conditions, this snapshot is taken immediately after formation of the first protostar. The three TreeCol--based methods refer to the different implementations of this algorithm (section \ref{sec:coolingmethods}) and the Sobolev and Gnedin method are only based on local quantities. For the relevant, optically thick densities above $\sim 10^{10}\ccm$, the local methods generally yield higher values than the more accurate TreeCol methods. See the electronic edition of the Journal for a colour version of this figure.
}
\label{plots_rho_col_syn}
\end{figure}

Before we apply our new method to cosmological halos, we test its accuracy in a simple test scenario. We start from a spherical, primordial cloud with a mass of $10^3M_\odot$, an initial density of $\sim 10^4\ccm$ and a temperature of $\sim 250$K. The gas is represented by $64^3$ SPH particles and we follow the simulation to densities above $\sim 10^{13}\ccm$, where we insert sink particles. For each different method, we run the simulation independently and compare the effective \htwo column densities in Figure \ref{plots_rho_col_syn}.
The lookup approach explicitly accounts for the true line overlaps and therefore is the most exact method. While ``TreeCol corrected Sobolev'' accounts for the correction factor and reproduces ``TreeCol lookup'' remarkably well, ``TreeCol Sobolev'' corresponds to the uncorrected Sobolev method and generally underestimates the effective column density. Consequently, this correction factor is mandatory to account for the exact line overlaps.

Furthermore, the commonly used Sobolev method and the Gnedin approach, which are only based on local gas quantities, overestimate the column density in the optically thick regime above $\sim 10^{10}\ccm$. These two local methods yield almost the same results for the effective column density, which means that the relation $v_\mathrm{th}/|\mathbf \nabla \cdot {\mathbf v}| \approx n_{\mhtwo}/|\mathbf{\nabla} n_{\mhtwo}|$ holds for the spherical collapse scenario. This illustrates that the density is generally not constant within the Sobolev length, but rather varies on the same length scale as the velocity field. Hence, even in this simple test case, the local methods are not able to determine the column density properly.

\subsection{Comparison of Cooling Approaches}
\label{section_comparecooling}
In this section, we test our approaches in more realistic cosmological halos. We first use the TreeCol--based simulations and determine all relevant information by post--processing these output files. By doing so, we can focus on the intrinsic differences of the methods (determined under the same physical conditions) rather than comparing different simulations with presumably different dynamics. The plots in this section represent averages over all four cosmological halos.

\begin{figure}[t]
\centering
\includegraphics[angle=-90]{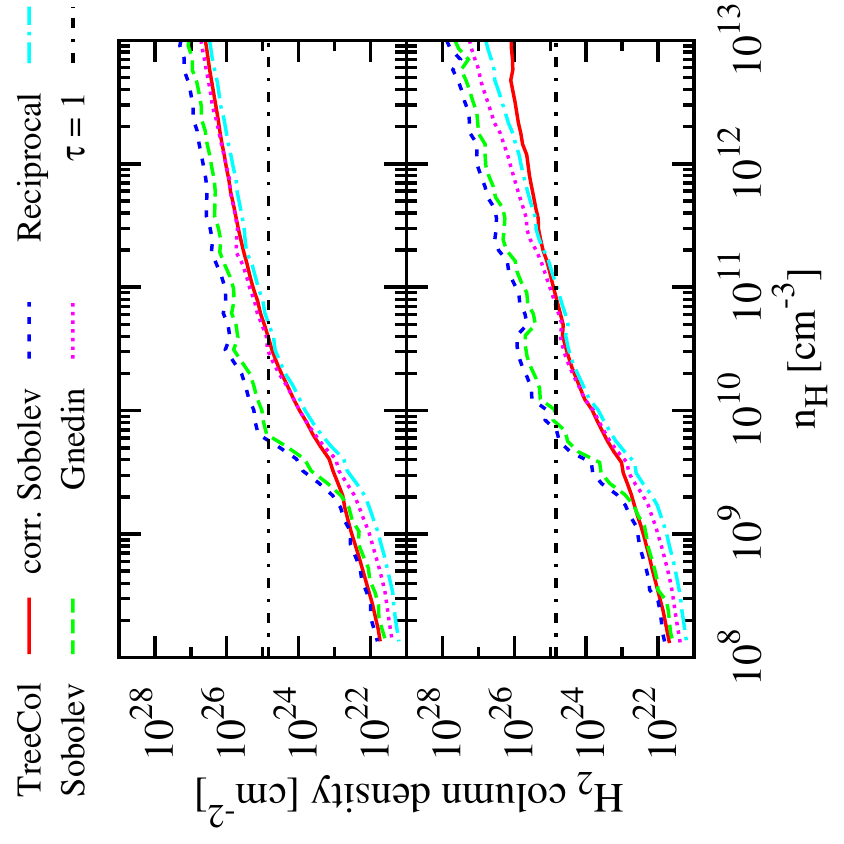}
\caption{Effective \htwo column density as a function of $n_\mathrm{H}$, the number density of H nuclei, for different cooling approaches. These snapshots are taken immediately before formation of the first sink particle (top) and immediately after formation of the second sink particle (bottom). ``Reciprocal'' represents the results based on the reciprocal sum of the Gnedin and the corrected Sobolev length. The solid red line indicates the TreeCol approach, which should be the most accurate of our models and the black dot-dashed line represents the \htwo column density above which the gas is optically thick, computed assuming a fixed temperature $T=1000$\, K. Almost all methods tend to overestimate the effective column density in the optically thick regime, some by up to two orders of magnitude.
The TreeCol column density decreases with time, while the other methods yield even higher column densities in the high--density regime. See the electronic edition of the Journal for a colour version of this figure.}
\label{plots_rho_col}
\end{figure}

We want to find an accurate cooling approach that reproduces the TreeCol method best and hence, we compare the commonly used Sobolev approximation with the corrected Sobolev approximation, the Gnedin approach and our new reciprocal method, which combines the corrected Sobolev and Gnedin approach. The column density for these different approaches can be seen in Figure \ref{plots_rho_col}.
Generally, the local methods overestimate the effective column density in the optically thick regime ($n \gtrsim 10^{10} \ccm$). Especially at later stages of the collapse these differences increase, because the slope of the photon escape fraction as a function of density flattens with time in the optically thick regime for the TreeCol method. The (corrected) Sobolev method overestimates the column density all the time, whereas the Gnedin and reciprocal approaches overestimate the column density only for high densities. The latter yield accurate fits for the density regime $10^9 \ccm \lesssim n \lesssim 10^{12} \ccm$, whereby one should keep in mind that the relative importance of \htwo line cooling decreases above $n \simeq 10^{13} \ccm$.
Realising that the Sobolev approximation already overestimates column densities, one might ask why we need this additional correction factor, which makes the approximation even worse. A detailed answer to this question is given in section \ref{sec:SobFails}, but we already want to emphasise the total neglect of any density gradient. The Sobolev approximation assumes a constant density, although the density of molecular hydrogen generally decreases when moving radially outwards. Hence, this approximation is not valid, but leads to an overestimation of the column density (already for the uncorrected Sobolev method). 

\begin{figure}[t]
\centering
\includegraphics[angle=-90]{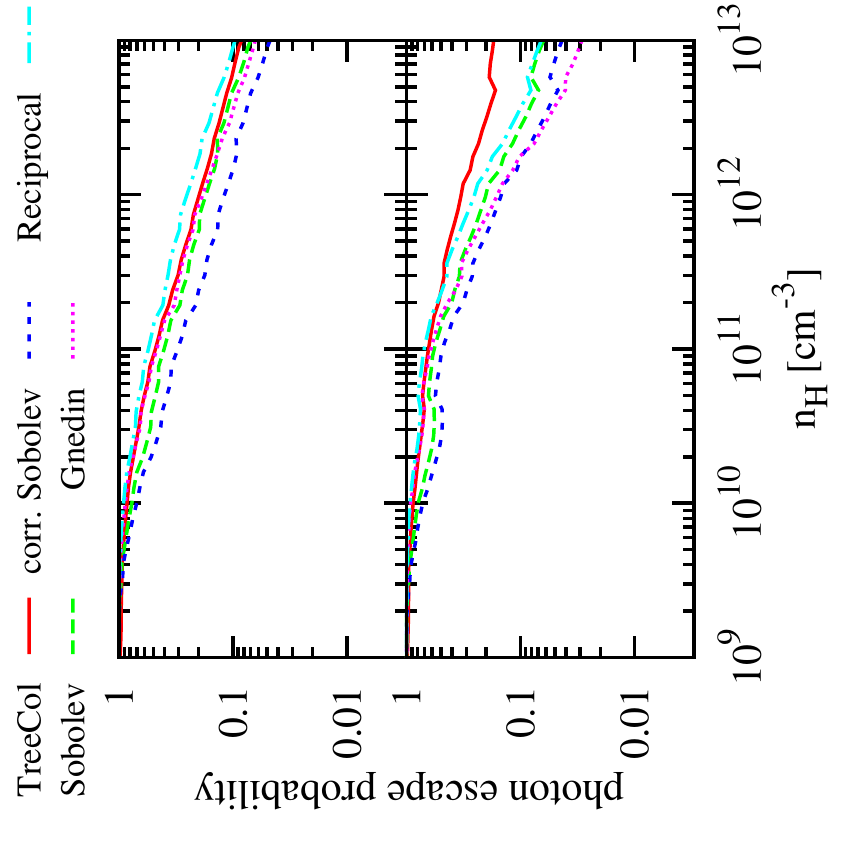}
\caption{Photon escape probability as a function of density for different cooling approaches. These snapshots are taken immediately before formation of the first sink particle (top) and immediately after formation of the second sink particle (bottom). The solid red line indicates the TreeCol approach, which should be fitted by the other methods.
While all the methods provide a good fit to the TreeCol method at low densities, the slope of the TreeCol approach flattens for later stages and the other methods cannot reproduce this behaviour. See the electronic edition of the Journal for a colour version of this figure.}
\label{plots_rho_opac_methods}
\end{figure}

The photon escape probability as a function of density can be seen in Figure \ref{plots_rho_opac_methods} for different methods.
Although there are slight differences between the individual approaches, all methods seem to agree well with the TreeCol approach prior to the formation of the first sink particles. However, at later times, the TreeCol approach yields higher values for the photon escape probability, corresponding to a smaller effective opacity of the cloud. The other methods, which depend only on local quantities, cannot reproduce this behaviour and therefore underestimate the photon escape probability.

\begin{figure}[t]
\centering
\includegraphics[angle=-90]{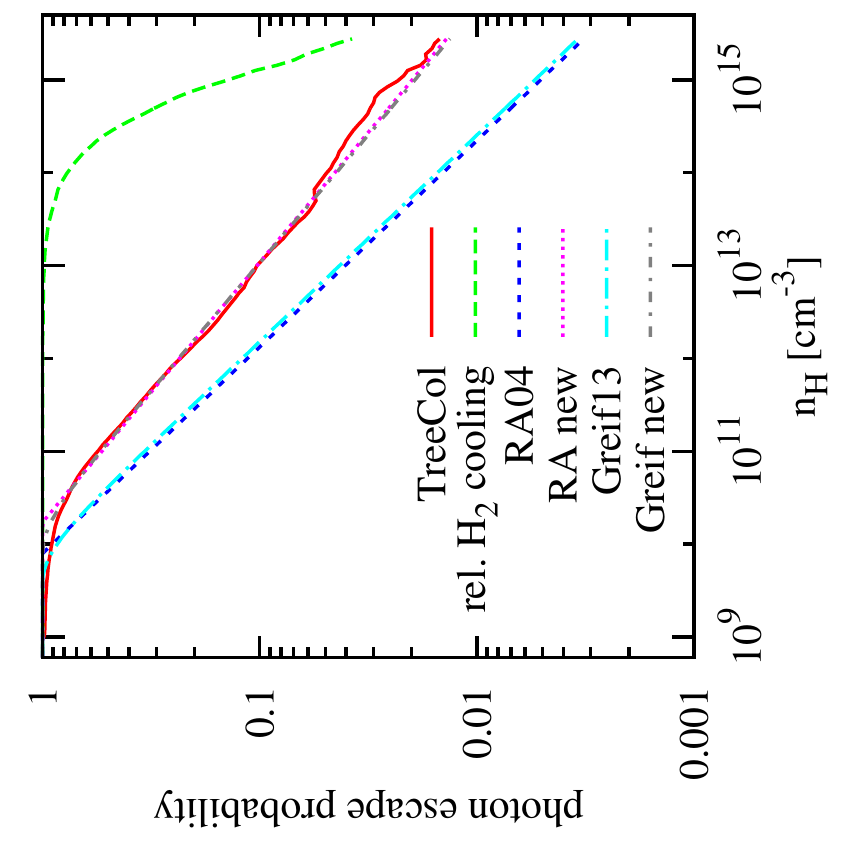}
\caption{Photon escape probability as a function of density for two analytical fitting functions (the green dashed line shows the relative strength of \htwo cooling). The snapshots are taken before formation of the first sink particle. The solid red line indicates the TreeCol approach, which should be fitted by the analytic formulae. The blue lines represent the original fits by \citet{ra04} and \citet{gsb13}, whereas the purple lines are their fit functions with the parameters adjusted to match our TreeCol method. Although the original fits are not reproduced, the data can generally be fitted by an analytic function for each specific time. See the electronic edition of the Journal for a colour version of this figure.}
\label{plots_rho_opac_analyt}
\end{figure}

\begin{figure}[t]
\centering
\includegraphics[angle=-90]{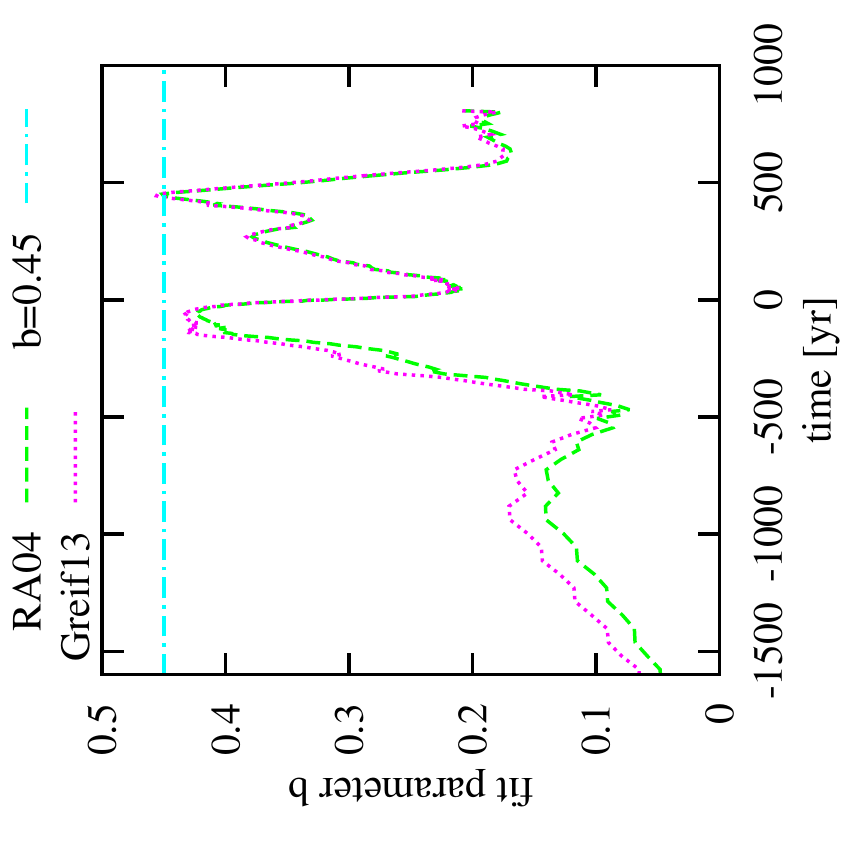}
\caption{Time evolution of the fit parameters $b_\mathrm{RA}$ (equation \ref{RA_fit_formula}) and $b_\mathrm{G}$ (equation \ref{Greif_fit_formula}) to the TreeCol data (formation of the first sink particle at $t=0$, averaged over four halos). The blue dash--dotted line illustrates the original slope of the fits ($b=0.45$). Generally the newly fitted slopes are shallower than the originally proposed value. However, since variations in the slope are of chaotic nature, it is impossible to find one global fit function. See the electronic edition of the Journal for a colour version of this figure.}
\label{opacity_fit}
\end{figure}

\begin{figure}[t]
\centering
\includegraphics[angle=-90]{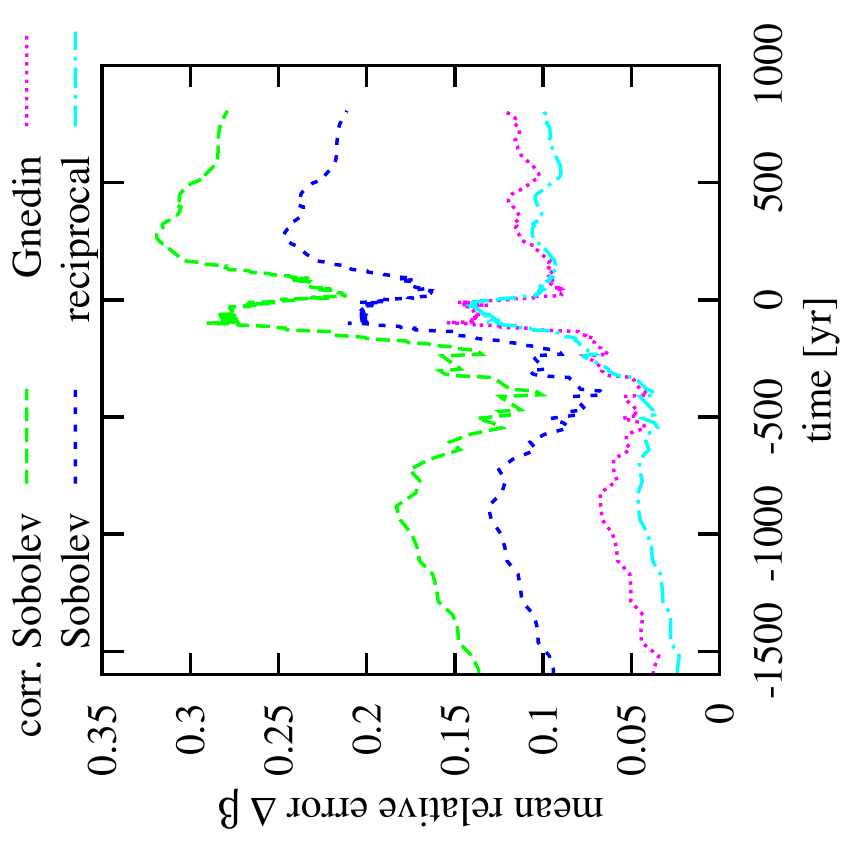}
\caption{Mean relative error of the photon escape probability for different methods as a function of time (averaged over four halos, at $t=0$ the first sink particle forms) for all particles above $n > 10^9 \ccm$. At early times, all photon escape probabilities are close to one and thus their mean error is small. At later stages, however, the relative error is between $5-30\%$ and the accuracy of the individual methods differ significantly. Data from all four simulations are combined in this plot. See the electronic edition of the Journal for a colour version of this figure.}
\label{opacity_error_1e9}
\end{figure}

We compare the analytical fitting formulae to the TreeCol method by using the parametrised formulae (equation \ref{RA_fit_formula} and \ref{Greif_fit_formula}) with the fit parameters $n_\mathrm{RA}$, $b_\mathrm{RA}$, $n_\mathrm{G}$, and $b_\mathrm{G}$. In Figure \ref{plots_rho_opac_analyt} we see the original functions and the best fits to the TreeCol data.
The newly adjusted fits minimise the weighted scatter sum
\begin{equation}
 \sum _{i=1} ^{N_\mathrm{SPH}} \frac{|\beta _\mathrm{analytic} (i) - \beta _\mathrm{TreeCol} (i)|}{\beta _\mathrm{TreeCol} (i)} \frac{\Lambda _{\mathrm{H}_2} (i)}{\Lambda _{\mathrm{H}_2} (i) + \Lambda _\mathrm{CIE}(i)},
\end{equation}
where $N_\mathrm{SPH}$ is the number of particles, $\beta _\mathrm{analytic} (i)$ is the analytic fit, $\beta _\mathrm{TreeCol} (i)$ is the photon escape probability based on the TreeCol approach, $\Lambda _{\mathrm{H}_2} (i)$ is the \htwo cooling rate, and $\Lambda _\mathrm{CIE}(i)$ is the CIE cooling rate for the $i$--th particle, respectively. The weighting by the relative cooling rate accounts for the decreasing relevance of \htwo cooling at higher densities.\\
For each single snapshot, the exact data can be fitted remarkable well by an analytic formula, but we note that both fit parameters vary strongly during the collapse. Although the original fits are a satisfying approximation at the beginning of the collapse, their accuracy decreases during the collapse and they totally miss the true photon escape probabilities at later stages.\\
The time evolution of the fit parameter $b$ (equation \ref{RA_fit_formula} and \ref{Greif_fit_formula}) reveals an interesting insight into the structure of the collapse. In Figure \ref{opacity_fit} we compare the evolution of $b$ for both fit formulae with time.
It quantifies the slope and therefore represents the dependence of the opacity on density. The parameters $b$ and $n_0$ are fitted simultaneously, but the plot only contains the slope because it reveals more information about the underlying physics. For large $b$, the cloud becomes opaque with increasing density very promptly, whereas for a shallow slope the opacity remains low although the density increases. The blue dash--dotted line illustrates the original slope for both fits of $b=0.45$,  but the newly fitted formulae to the TreeCol data reveal that the actual slope is shallower and most notably varies with time. The small value of $b$ for $t<-500$yr is a numerical artefact, because at these times there are almost no particles in the optically thick regime, which might define a distinct slope. On the other hand, the shallow slope of $0.2 \leq b \leq 0.4$ during the later stages of the collapse can be related to the flattening of the cloud.
Consequently, it is impossible to find one single analytic fit, which describes the dynamics of the collapse completely. However, the mean parameters that fit the TreeCol data best from $500$yr before formation of the first protostar until the end of our simulations are $n_\mathrm{G} = 5 \times 10^9 \ccm$ and $b_{\mathrm{G}}=0.32$.\\
In order to analyse the accuracy of the different cooling approaches quantitatively, we determine the relative error of the photon escape probability for each method. Therefore, we compare the photon escape probabilities of all particles above a certain density threshold to the ones determined with the TreeCol method. The density threshold is necessary, because below $n = 10^9 \ccm$, the escape probability is very close to one anyway and consequently there are no significant deviations between the methods. The relative error $ |\beta _\mathrm{method} - \beta _\mathrm{TreeCol}| / \beta _\mathrm{TreeCol}$ is weighted by the relative \htwo cooling rate, so that the mean relative error of the photon escape probability can be expressed as
\begin{align}
 \overline{\Delta \beta} = \frac{1}{N_\mathrm{thresh}} &\sum _{i=1} ^{N_\mathrm{thresh}} \frac{|\beta _\mathrm{method} (i) - \beta _\mathrm{TreeCol} (i)|}{\beta _\mathrm{TreeCol} (i)} \nonumber \\
 &\times \frac{\Lambda _{\mathrm{H}_2} (i)}{\Lambda _{\mathrm{H}_2} (i) + \Lambda _\mathrm{CIE}(i)},
\end{align}
where $N_\mathrm{thresh}$ is the number of particles above a certain density threshold. We chose the density threshold $10^9 \ccm$ to analyse how accurate the methods are in this density regime. The time evolution of the mean relative photon escape probability error for this threshold can be seen in Figure \ref{opacity_error_1e9}.
The relative error is small at early times in the collapse because all photon escape probabilities are close to one, whereas at later times, more particles enter the optically thick regime and the relative errors rise to values between $5-30\%$. The reciprocal approach is slightly better than the Gnedin approximation, although both yield errors between $5-15\%$ throughout the simulations. Interestingly, the corrected Sobolev approximation yields higher errors ($15-30\%$), whereas the uncorrected one yields errors between $10-25\%$. This behaviour is caused by the fact that the Sobolev approximation already overestimates column densities because it neglects the density gradient. Since the reciprocal method is mostly dominated by the Gnedin approximation, we can conclude that the effective column density is more strongly influenced by the density gradient than by the relative velocities. However, the velocity information still improves the fit and is therefore important for a 
proper 
treatment.
The original analytic fits are comparatively inaccurate with errors between $20-50\%$ and are therefore not displayed in this figure. However, the analytical fitting function proposed by \citet{gsb13} yields slightly better results than the one proposed by \citet{ra04}.
The same analysis for higher density thresholds ($10^{10} \ccm, 10^{11} \ccm$) shows qualitatively similar results, but with a trend to higher mean relative errors.

\section{Collapse and Fragmentation}
\label{sec:frag}
Here, we analyse the fragmentation behaviour of the gas with the different cooling implementations. We first compare the number of protostars that form in each simulation, analyse the dynamics of the collapse and then determine the susceptibility to fragmentation for the Sobolev method and the TreeCol approach.

\subsection{Number of Protostars}
We are interested in how the treatment of the cooling affects the characteristic mass range for Pop III stars. Based on our simulations, we compare the number of sink particles in each halo at the end of the simulations. However, due to a prohibitively small chemical time step and a too simplified treatment of feedback, we end the simulations before accretion has terminated. In order to compare the number of sink particles, we choose for each halo the snapshots with the same amount of accreted mass onto the sink particles. This condition enables us to compare the clouds for the Sobolev--based and the TreeCol--based runs at the same stage of the accretion process. The resulting numbers of sink particles are given in Table \ref{table:sinks}.

\begin{deluxetable}{cccc}
\tablecaption{Number of sink particles per halo}
\tablehead{\colhead{Halo} & \colhead{$m_\mathrm{acc,tot}$} & \colhead{Sinks Sobolev} & \colhead{Sinks TreeCol} \\ 
\colhead{} & \colhead{($M_\odot$)} & \colhead{} & \colhead{} } 

\startdata
1 & 11.0 & 1 & 2 \\
2 & 5.1 & 2 & 3 \\
3 & 3.3 & 1 & 3 \\
4 & 10.3 & 5 & 4
\enddata

\tablecomments{Comparison of the number of sink particles for 4 halos between the Sobolev approximation and the TreeCol approach. The same amount of accreted stellar mass guarantees comparability between the methods. In all but one simulation the TreeCol approach yields more sink particles.}
\label{table:sinks}
\end{deluxetable}

\begin{figure}[t]
\centering
\centerline{\includegraphics[angle=-90]{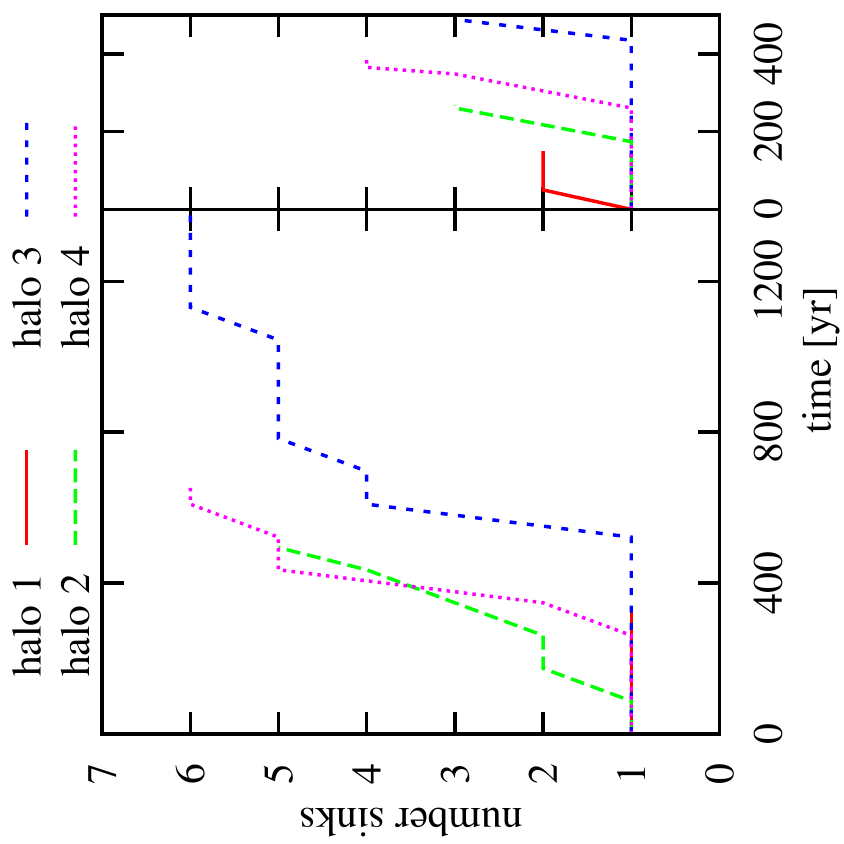}}
\caption{Number of sink particles as a function of time for the Sobolev--based runs (left) and the TreeCol--based runs (right). The clouds in the TreeCol approach fragment earlier, yielding more than two sink particles in each run, whereas the Sobolev approximation yields one to six sink particles per halo. See the electronic edition of the Journal for a colour version of this figure.}
\label{plot:timeN}
\end{figure}

The TreeCol approach yields more sink particles in all but one halo. However, the time evolution of the number of sink particles (see Figure \ref{plot:timeN}) suggests that this effect might be caused by a delay of the fragmentation in the Sobolev--based runs.
While the Sobolev approximation causes the disc to fragment into 1-6 protostars, the TreeCol approach causes slightly earlier fragmentation with at least two protostars in each halo. Since we had to end the simulations at this point, the final number of protostars cannot be fully constrained in this case. However, the TreeCol method accelerates the collapse and the gas fragments earlier compared to the Sobolev approximation. The accretion rates onto individuals protostars are in the range $3.3 \times 10^{-4} M_\sun \mathrm{yr}^{-1} \leq \dot{M} \leq 1.2 \times 10^{-2} M_\sun \mathrm{yr}^{-1}$ for the Sobolev approximation and $1.2 \times 10^{-3} M_\sun \mathrm{yr}^{-1} \leq \dot{M} \leq 1.7 \times 10^{-2} M_\sun \mathrm{yr}^{-1}$ for the TreeCol approach in all four halos.

\subsection{Collapse Dynamics}
\begin{figure*}[t]
\centering
\includegraphics[width=14cm]{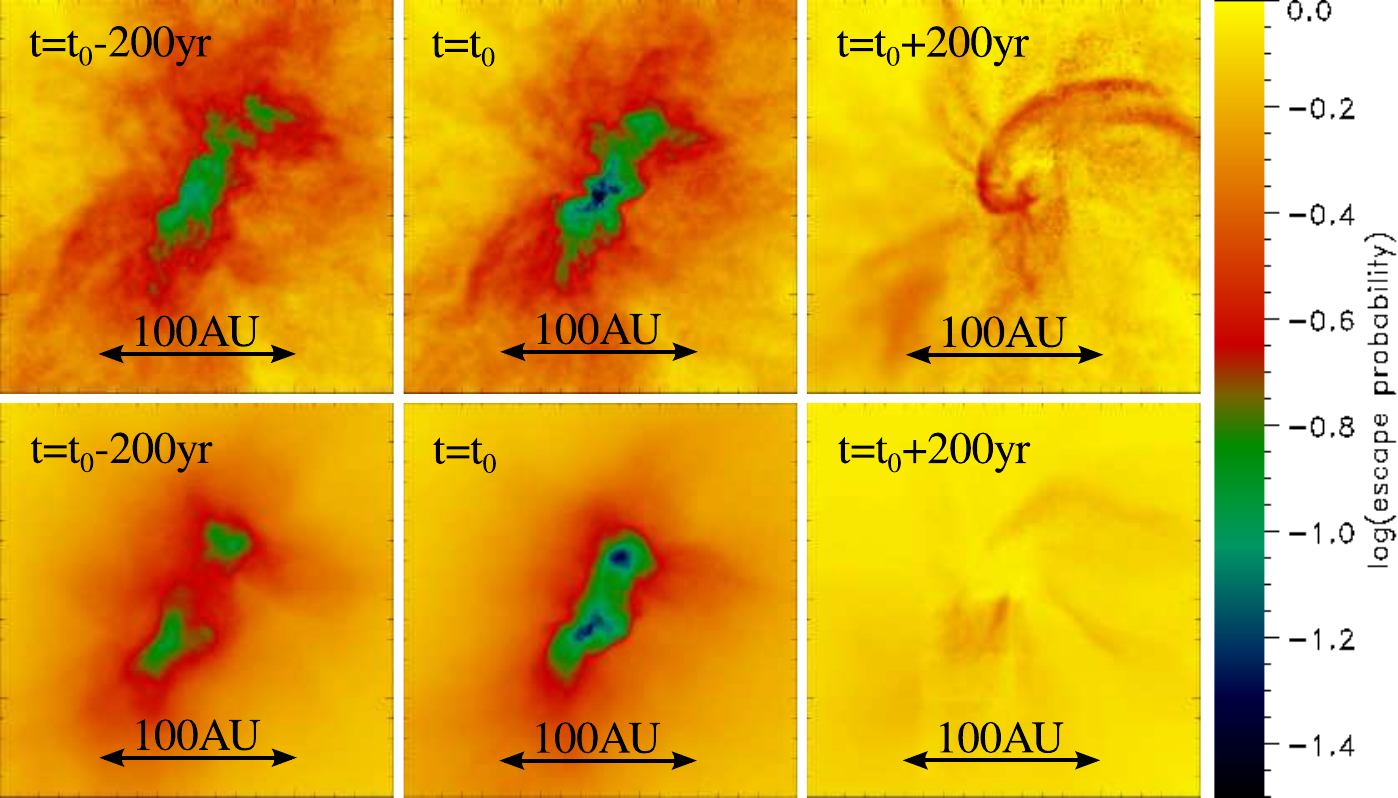}
\caption{Mass--weighted projections of the photon escape probability in the central region for the Sobolev method (top) and the TreeCol method (bottom) at different times during the collapse: before sink formation (left), immediately after sink formation (middle), and $\sim 200$yr after the first sink particle formed (right). The photon escape probability appears to be smoother for the TreeCol approach and especially at late stages of the collapse the photon escape probability is close to unity for the TreeCol--based run. This increase in the photon escape probability is not reproduced by the Sobolev approximation. These plots are based on halo 1 but are representative for the other realisations. See the electronic edition of the Journal for a colour version of this figure.}
\label{plot_compare_opac}
\end{figure*}

\begin{figure}[t]
\centering
\centerline{\includegraphics[angle=-90]{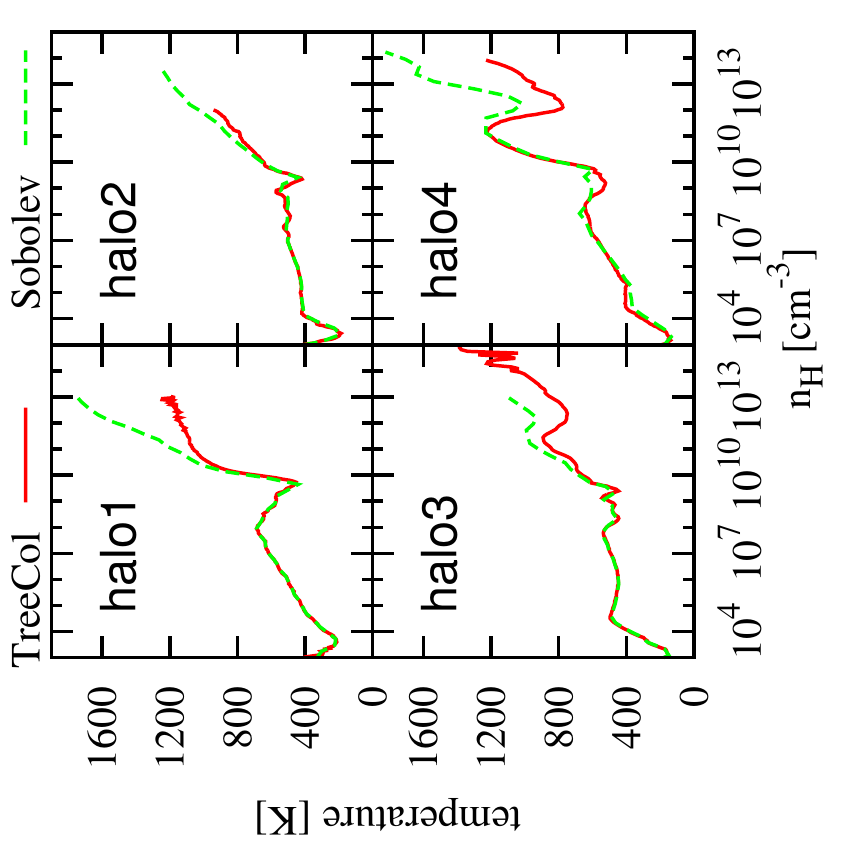}}
\caption{Temperature as a function of density immediately after formation of the first sink particle for the four halos. Comparison of the Sobolev approximation (green dashed) and the TreeCol method (red solid). In the inner, high--density regime, the TreeCol approach yields lower temperatures by up to $\sim 500$K. See the electronic edition of the Journal for a colour version of this figure.}
\label{plot_compare_temps}
\end{figure}

In order to study the fragmentation behaviour independently of the number of sinks, we first analyse the dynamics of the collapse. This analysis is mainly based on the fact that the temperature profile of a gas cloud directly influence the collapse to protostars by affecting the local Jeans mass, which is the absolute minimum requirement for collapse and is a strong function of temperature \citep{cgsgkb11}. Hence, we compare the basic quantities between the four runs using the TreeCol approach and the four runs using the Sobolev approximation. The first quantity to compare is naturally the photon escape probability. Figure \ref{plot_compare_opac} illustrates the time evolution of the photon escape probability for both methods.
While the spatial distribution of the photon escape probability seems to be smoother for the TreeCol--based run, the distribution is comparatively structured for the Sobolev approximation. Additionally, we see another important difference. Using the TreeCol method yields values for the photon escape probability close to one at the end of the simulations, whereas the Sobolev approximation yields smaller photon escape probabilities. Obviously, the Sobolev approximation is not able to capture the flattening of the cloud (see section \ref{sec:SobFails}).
Furthermore, the Sobolev--based simulation develops only one central core, while an elongated core with two peaks is formed in the TreeCol--based run. Since the photon escape probability has a direct influence on the cooling rates, we are consequently interested in the \htwo cooling rate. This cooling rate is not significantly higher for the TreeCol--based run, although the photon escape probabilities are higher. This can be explained by the cooling implementation: due to the presumably higher cooling rate in the TreeCol--based run, thermal equilibrium is reached for lower temperatures. Thus, the cloud simply remains cooler instead of increasing its cooling rate significantly.
Consequently, the gas in the Sobolev--based runs is generally hotter, which can also be seen in Figure \ref{plot_compare_temps}.
The difference of temperatures in this inner regime can be up to $\sim 500$K, which influences the fragmentation behaviour significantly.

\subsection{Stability Analysis}
Finally, we compare different fragmentation criteria of the clouds. The possibility of fragmentation in primordial clouds has been considered by many authors \citep[e.g.][]{sy77, cgk08, tao09, sgb10, cgkb11, getal11, cgsgkb11, getal12, md13, gsb13, b13,sb14}. Fragmentation is a very chaotic, non--deterministic process and the actual outcome depends sensitively on the initial conditions \citep{gfbk11,gfbk12a,gfbk12b}. Nevertheless, there are three analytic expressions that help to quantify the possibility of a gas cloud fragmenting:
\begin{itemize}
 \item In order to locally contract instead of globally collapse, a necessary criterion for fragmentation of a gas cloud is \citep{ro77}
\begin{align}
 \frac{t_\mathrm{cool}}{t_\mathrm{ff}} < 1,
\end{align}
 where $t_\mathrm{cool} = n k_B T/(\Lambda [\gamma -1])$ is the cooling time with the net cooling rate $\Lambda$ and $t_\mathrm{ff} = \sqrt{3 \pi/(32 G \rho)}$ is the free--fall time.
 \item \citet{t64} analysed the stability of rotating gas discs and derives the instability criteria
\begin{align}
 Q=\frac{c_s \kappa}{\pi G \Sigma} <1,
\label{eq_toomre}
\end{align}
where $\kappa$ is the epicyclic frequency and $\Sigma$ is the surface density of the disc. Formally, this criterion is only valid for thin discs whereas \citet{gl65} extended the criterion by requiring $Q<0.676$ for a finite--thickness isothermal disc to fragment.
\item \citet{g01} investigated the nonlinear outcome of a stability analysis of a Keplerian accretion disc. Based on numerical experiments, he derived the instability criterion
\begin{align}
 \frac{t_\mathrm{cool}}{3 \omega ^{-1}} < 1,
 \label{eq_gammie}
\end{align}
where $\omega$ is the orbital frequency. The Gammie criterion expresses the possibility that pieces of the disc cool and collapse before they have the opportunity to collide with one another in order to reheat the disc.
\end{itemize}
In their study on the formation and evolution of primordial protostellar systems, \citet{getal12} find the Toomre criterion insufficient for the quantification of gravitational instability and additionally use the Gammie criterion. Non of these criteria guarantees fragmentation \citep{yoha06}, but a combination of them yields a reliable quantification of instabilities in the gas disc.\\

\begin{figure*}[t]
\centering
\includegraphics[width=0.9\textwidth]{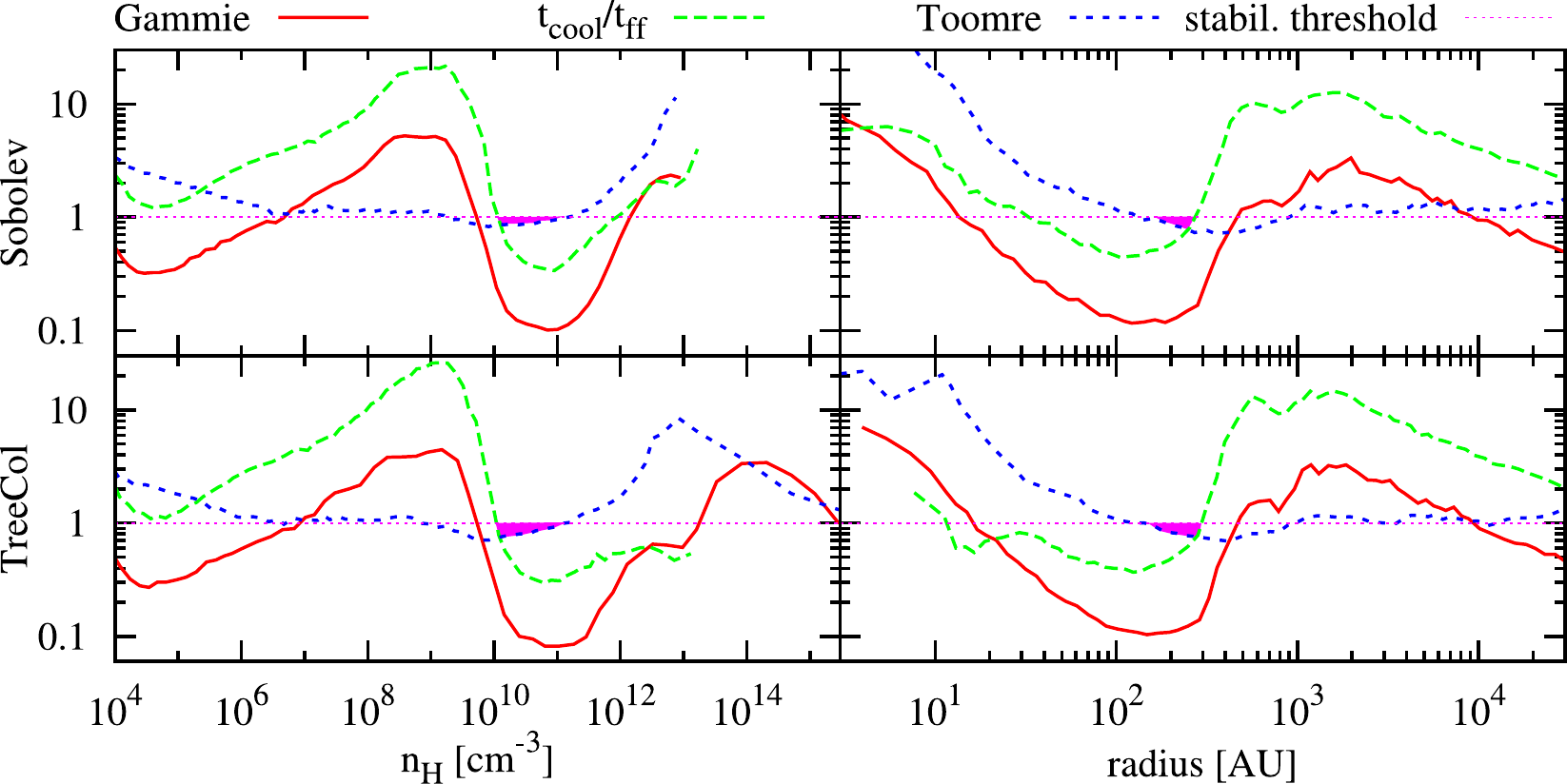}
\caption{Three different fragmentation criteria as a function of density (left) and radius (right) for the Sobolev--based (top) and the TreeCol--based run (bottom). The data are averaged over 4 halos. Since these plots are qualitatively equal for all four halos, the average yields a representative quantification of fragmentation in primordial gas. These snapshots are taken between formation of the first and second protostar and quantify the susceptibility to disc fragmentation. We expect the discs to fragment if all three criteria are below one (purple dotted line). Although the requirement that all three fragmentation criteria have to be fulfilled simultaneously is a very strict one, all simulations are likely to fragment in the density regime $10^{10} \ccm \leq n \leq 10^{11}\ccm$ at a radius of $\sim200$AU from the densest region (purple area). Since the temperature in this density regime is hardly affected by the choice of the cooling approach, this fragmentation 
regime is only slightly pronounced in the TreeCol--based 
run. However, at 
higher densities, $t_\mathrm{cool}/t_{ff}$ remains longer in the instability 
regime and hence we expect the TreeCol--based runs to be more susceptible to 
fragmentation in the high--density regime, because the gas can locally contract instead of globally collapse. Because of our sink accretion radius of 10AU, the central region is artificially stabilised. See the electronic edition of the Journal for a colour version of this figure.}
\label{fig:fragmentation}
\end{figure*}

We compare the radial and density profile of these fragmentation criteria for the Sobolev--based and the TreeCol--based runs. We do this directly after formation of the first protostar to ensure comparability between the different runs and methods (Fig. \ref{fig:fragmentation}).
The requirement that all three fragmentation criteria have to be fulfilled simultaneously is very strict, because each criterion alone already quantifies stability. Applying this conservative criterion shows that all simulations are susceptible to fragmentations in the density regime $10^{10} \ccm \leq n \leq 10^{11}\ccm$ at a radius of $\sim200$AU from the densest region. However, the Gammie criterion seems to be a less restrictive version of the Rees--Ostriker--criterion and consequently yields no additional constraint to this regime, because it is fulfilled anyway. The regime of fragmentation and the instability of the gas cloud is slightly higher for the TreeCol approach. Especially the cooling time criterion remains significantly longer under the stability threshold for the TreeCol--based run, which is mainly a consequence of the lower temperatures, as discussed before. Consequently, the high--density gas is more susceptible to fragmentation in the TreeCol--based runs. Since these are 
averaged profiles over four 
halos, we expect these results to be representative for primordial star formation.

\section{DISCUSSION}
\label{sec:discussion}
In this section, we discuss the previous results, point out shortcomings of commonly used methods, and comment on the expected different fragmentation behaviour.

\subsection{Cooling Approaches}
\label{sec:SobFails}
\begin{figure*}[t]
\centering
\includegraphics{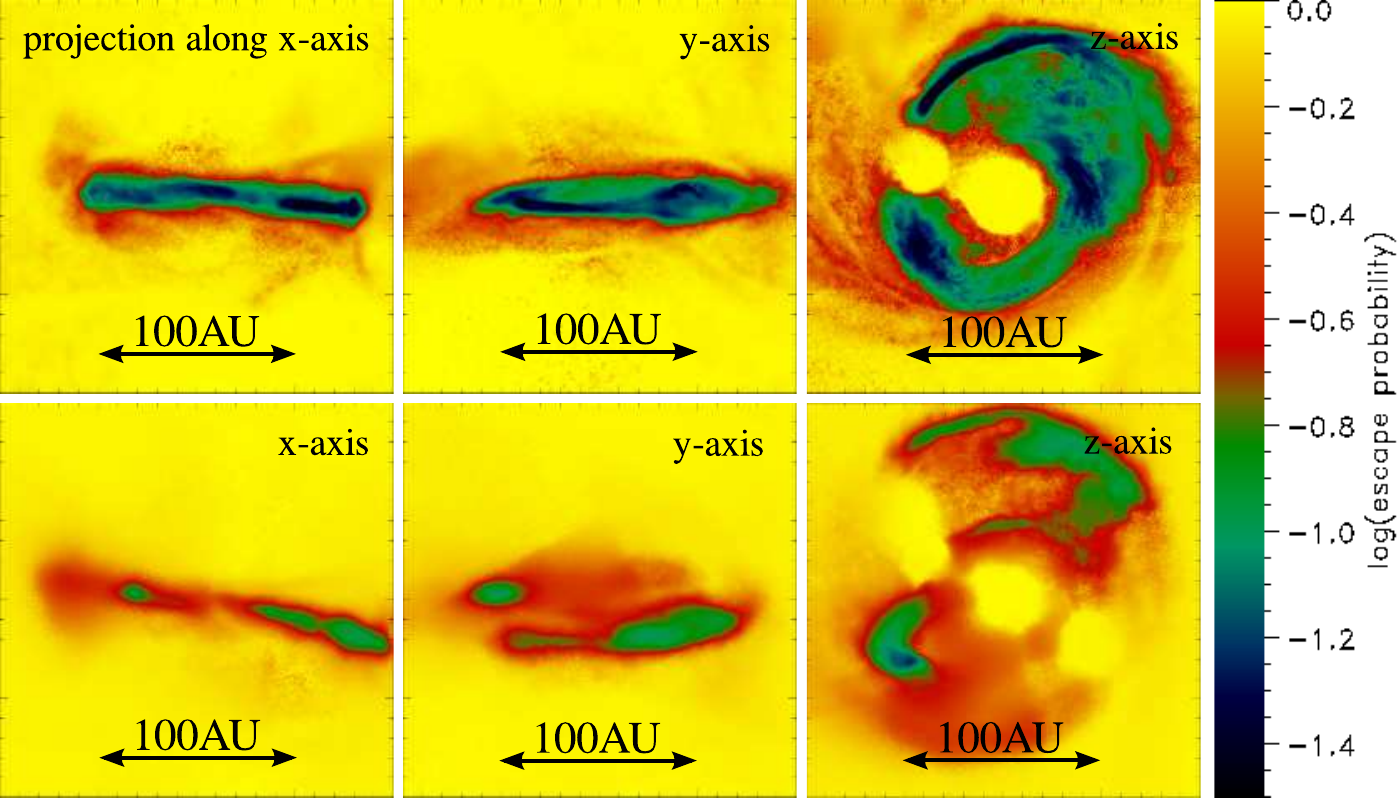}
\caption{Mass--weighted projections of the photon escape probability along the x--(left), y--(middle), and z--axis (right) for the Sobolev approximation (top) and the TreeCol method (bottom). This disc of halo 4, after several protostars have formed, is exemplary for the occurring processes. The flattened structure enables the photons to escape perpendicular to the disc. However, this enhanced photon escape probability is only reproduced by the TreeCol method, because the Sobolev approximation is based on the assumption of spherical symmetry. See the electronic edition of the Journal for a colour version of this figure.}
\label{plot_xyzopac}
\end{figure*}

The local and isotropic approximations fail in determining the proper effective column densities for \htwo cooling. Although they yield acceptable results up to the formation of the first protostars, most approaches break down during later phases of the collapse, because they cannot capture the evolving dynamical structure, particularly once a flattened accretion disc has formed. The appropriate consideration of the overlap of spectral lines is very important, although even the uncorrected Sobolev approximation generally overestimates column densities. Further shortcomings are mostly related to the assumption that all relevant quantities are constant within the characteristic length scale. This simplification avoids the evaluation of integrals along the line of sight but is formally only valid for large velocity gradients.
The analysis of the relevant quantities shows that the \htwo number density, thermal velocity and the gradient of the velocity vary by up to two orders of magnitude within one Sobolev length. While the thermal velocity seems to be rather constant, the \htwo number density varies most strongly. Since the distribution around the central value is not symmetric, these effects do not cancel out and generally lead to an overestimation of column densities.\\
Due to angular momentum conservation in the infalling material, a disc forms around the first protostar and the photon escape probability is enhanced perpendicular to the disc. An isotropic column density estimation that depends only on local quantities, can not capture this feature and therefore generally underestimates the photon escape probability (see Figure \ref{plot_xyzopac}).\\
The Sobolev approximation is not able to capture this angular dependence and consequently underpredicts the angle--averaged photon escape probability. Comparing the escape probabilities in different directions for one and the same particle quantitatively yields high angle--dependent variations. Shortly before formation of the first sink particle, the photon escape probabilities in the inner $\sim100$AU vary by factors of up to$10-100$ in different HEALPix pixels. During the further collapse, this value decreases but is still between $2-20$ on average in the centre of the cloud. This clearly demonstrates the failure of the local approximation in modelling inhomogeneous, non--spherical density distributions. \citet{hy13} also mention the problem of direction--dependent escape probabilities during their comparison of different cooling implementations in simulations of primordial star formation.
Moreover, the Sobolev approximation is highly sensitive to small-scale details of the velocity field, and hence improving the resolution and accuracy with which one models the sub-sonic turbulence in the gas actually makes the method worse, because locally you start to see more of the turbulent velocity gradients, but these persist only over short length scales, not globally \citep{g14a}.

Although the TreeCol--based method is computationally expensive, it is able to capture the non--uniformity of the \htwo distribution in a way that local approximations cannot manage. It is also significantly cheaper than the more robust method employed by \citet{g14a}, although a comparison between the accuracy of the method proposed here and that by \citet{g14a} still needs to be done.

Regarding the local methods, there is no general recommendation. Most of these methods assume spherical symmetry and hence their validity breaks down when a disc forms around the first protostar. Since the accretion process through the disc lasts for several thousand years or more, one should use an approach which yields a suitable long--time accuracy.
The reciprocal method is the only approach that considers both gradients in density and the Doppler--shifting of spectral lines. Therefore, this method is the most accurate approach for the determination of the effective column density even after subsequent sink particle formation. The difference between whether one uses the corrected or uncorrected Sobolev length for the determination of the reciprocal sum is small, but the corrected Sobolev length provides a more accurate method for the long--time evolution of the disc.
Analytic fits might be useful during the initial collapse of the primordial cloud, but also fail when the disc--like structure starts to form. Nevertheless, at any stage of the collapse, the escape probability as a function of mass can generally be fitted very accurately by an analytic formula, which could also be done on the fly during the simulation. Consequently, a hybrid version between the proper TreeCol--based determination of escape probabilities every several time steps and an analytic fit to these data for the steps in between should yield reliable results with less computational effort.

This new method for the determination of effective column densities can also be applied in other astrophysical scenarios. For example the photodissociation of \htwo in protogalaxies is a crucial process for the formation of supermassive black hole seeds \citep{letal13}. To model these processes properly, the effective column densities are needed to account for the effect of \htwo self-shielding against Lyman Werner photons \citep{whb11}. It can generally be applied in most scenarios of line transfer, where the radially infalling velocities of gas are higher than the turbulent gas velocities.

\subsection{Fragmentation}
\label{sec:frag2}
Fragmentation is a highly chaotic process and slight changes in the initial conditions or in the implementation of the governing physics can completely change the outcome. Therefore, looking just at the number of sink particles is not a valid quantification, especially since we do not capture the entire duration of the accretion and fragmentation process. Nevertheless, it is instructive to determine and compare different fragmentation criteria for the individual cooling approaches. We find that the TreeCol method seems to promote fragmentation. In other words, commonly used cooling approximations generally underestimate the number of Pop III stars. Moreover, the disc has just formed, when we end our simulations. Therefore, the differences between the individual cooling approaches and their effect on fragmentation might even be more pronounced for later stages of the collapse.
Since the overall mass accretion rates are roughly equal, regardless of the cooling implementation, we expect the Pop III stars in our TreeCol--based runs generally to have smaller masses than previous Sobolev--based studies have yielded.

\subsection{Caveats}
There are several open questions, shortcomings and approximations, which one should keep in mind, when interpreting the previously presented results. Since we follow the fragmentation of the cloud only for the first few hundred years after formation of the first protostar, we have no information about the physical conditions during the late stages of disc accretion. It is, however, likely that the disc--like structure will proceed to grow and most of our statements remain valid.\\
Furthermore, we do not account for mergers, which influence the final number of Pop III stars \citep{getal11, getal12, sb14}, nor do we include the effects of ionizing radiation. We determine radiative feedback under the assumption of a constant mass accretion rate and although the value of $10^{-2} M_\sun \mathrm{yr} ^{-1}$ seems to be justified by similar accretion rates in the simulations, a proper treatment is necessary. Besides, we should also keep in mind that the escape probability, as the theoretical basis for our cooling implementation, is an approximation by itself, e.g. we use one average escape probability for all photons, instead of determining the individual escape probabilities for each line separately. In this context, \citet{g14a} find that a multi--line, multi--frequency raytracing scheme does not alter these results significantly.
Magnetic fields might influence the collapse and act as a stabilising force. We have not included these effects in our simulation, but for a detailed discussion of the effects of magnetic fields in primordial star formation, see \citet{md13} and references therein.

\section{Conclusion}
\label{sec:conclusion}
We compared different implementations for the approximate treatment of optically--thick \htwo line cooling and analysed the fragmentation behaviour of primordial gas under these different methods. Since \htwo is the dominant coolant in primordial gas clouds, line cooling by molecular hydrogen is a crucial process to consider for the formation of Pop III stars. While the cooling rates in the optically thin regime can be calculated accurately, optically thick cooling is only poorly understood, although it has a strong influence on the temperature profile and fragmentation of the cloud. The commonly used Sobolev approximation has to be corrected for the effect of line overlap. However, since the Sobolev and other approximations for the effective column density assume isotropy and certain quantities to be constant on the relevant scales, they all fail if the system deviates strongly from spherical symmetry or if it has strong density gradients. While the cloud 
flattens and develops a disc during the collapse, the local approaches generally yield too small 
values for the photon escape probability (mean relative errors of $\sim 20\%$).
Existing analytical fitting formulae yield acceptable results up to the formation of the first protostar. Thereafter, the detailed functional form needs to be modified, because an analytical fitting formula derived at one time in the evolution will generally fail at other times, as the system is strongly out of dynamical equilibrium and therefore rapidly evolving.
Only the TreeCol--based methods are adaptively adjusting to the kinematic and morphological changes of the system. Capturing these dynamical features, the TreeCol--based methods yield lower temperatures in the centre of the cloud.\\
We find primordial gas is most susceptible to fragmentation in the density regime from $n=10^{10} \ccm$ to $n=10^{11} \ccm$. Whereas local methods lead to the formation of fewer and higher-mass Pop III stars, the TreeCol--based approach promotes a higher degree of fragmentation and therefore tend to result in the formation of more and lower--mass stars. Regardless of the cooling implementation, the protostars have very high mass accretion rates and the mass function will be dominated by high--mass stars. However, only high--resolution simulations with a proper treatment of \htwo cooling, magnetic fields, and feedback that run long enough to capture the entire accretion and fragmentation process can give a complete picture of Pop III star formation and the corresponding primordial IMF.

\begin{acknowledgements}
The authors acknowledge financial support from the Deutsche Forschungsgemeinschaft (DFG) via SFB 881 ``The Milky Way System'' (sub-projects B1, B2, and B8), and from the Baden-W\"{u}rttemberg-Stiftung by contract research via the programme Internationale Spitzenforschung II (grant P-LS-SPII/18). The simulations described in this paper were performed on the {\em kolob} cluster at the University of Heidelberg, which is funded in part by the DFG via Emmy-Noether grant BA 3706, and via a Frontier grant of Heidelberg University, sponsored by the German Excellence Initiative as well as the Baden-W\"urttemberg Foundation.
TH appreciates fruitful discussions with Clio Bertelli Motta and Volker Bromm. TH and RSK are supported from the European Research Council under the European Communitys Seventh Framework Programme (FP7/2007-2013) via the ERC Advanced Grant ``STARLIGHT: Formation of the First Stars'' (project number 339177). MS and TH acknowledges computing time from JUDGE super computing centre.
PCC is supported by grant CL 463/2-1, part of the DFG priority program 1573 ``Physics of the Interstellar Medium''.
MS, as part of IMPRS-HD, thanks financial support from Heidelberg Graduate School of Fundamental Physics (HGSFP) and wishes to thank Thomas Greif for help with performing numerical simulations.

\end{acknowledgements}

\bibliographystyle{apj}

\end{document}